\documentclass[sigconf,authorversion,nonacm]{acmart}

\usepackage[normalem]{ulem}
\usepackage{tikz}
\newcommand*\circled[1]{\tikz[baseline=(char.base)]{\node[shape=circle,draw,inner sep=2pt] (char) {#1};}}
\usepackage{graphicx}
\usepackage{amsmath}
\usepackage{graphics}
\usepackage{multirow}
\usepackage{algorithm}
\usepackage{algorithmic}

\usepackage{hyphenat}
\usepackage{graphicx}
\usepackage{amsmath}
\usepackage{geometry}
\usepackage{multirow}
\usepackage{subfig}
\usepackage{enumitem}

\definecolor{dong}{RGB}{0,0,0}

\AtBeginDocument{%
  }

\renewcommand\footnotetextcopyrightpermission[1]{}

\begin{document}


\newcommand{\name}{AttMemo\ }
\newcommand{\namenospace}{AttMemo}

\title{\name: Accelerating Self-Attention with Memoization on Big Memory Systems}



\author{Yuan Feng}
\affiliation{%
  \institution{University of California, Merced}
  \city{Merced}
  \country{USA}
}
\email{yfeng44@ucmerced.edu}

\author{Hyeran Jeon}
\affiliation{%
  \institution{University of California, Merced}
  \city{Merced}
  \country{USA}
}
\email{hjeon7@ucmerced.edu}

\author{Filip Blagojevic}
\affiliation{%
  \institution{Western Digital Research}
  \city{Milpitas}
  \country{USA}
}
\email{filip.blagojevic@wdc.com}

\author{Cyril Guyot}
\affiliation{%
  \institution{Western Digital Research}
  \city{Milpitas}
  \country{USA}
}
\email{cyril.guyot@wdc.com}

\author{Qing Li}
\affiliation{%
  \institution{Western Digital Research}
  \city{Milpitas}
  \country{USA}
}
\email{qing.li7@wdc.com}

\author{Dong Li}
\affiliation{%
  \institution{University of California, Merced}
  \city{Merced}
  \country{USA}
}
\email{dli35@ucmerced.edu}







\begin{abstract}
Transformer models gain popularity because of their superior inference accuracy and inference throughput. However, the transformer is computation-intensive, causing a long inference time. The existing works on transformer inference acceleration have limitations caused by either the modification of transformer architectures or the need of specialized hardware. In this paper, we identify the opportunities of using memoization to accelerate the self-attention mechanism in transformers without the above limitations. Built upon a unique observation that there is rich similarity in attention computation across inference sequences, we build a memoization database that leverages the emerging big memory system. We introduce 
a novel embedding technique to find semantically similar inputs to identify computation similarity. We also introduce a series of techniques such as memory mapping and selective memoization to avoid memory copy and unnecessary overhead. We enable 22\% inference-latency reduction on average (up to 68\%) with negligible loss in inference accuracy.
\end{abstract}

\setcopyright{none}
\settopmatter{printacmref=false} 
\maketitle

\section{Introduction}

Transformer models~\cite{devlin2018bert, he2020deberta, liu2019roberta, scao2022bloom, touvron2023llama, chatgpt} recently gain fame in various computing fields such as natural language processing (NLP) and computer vision (CV). Transformers provide superior inference  accuracy and throughput through parallelized self-attention mechanism~\cite{attention}. 
Different from the traditional models (e.g., recurrent neural networks (RNN)~\cite{rumelhart1986learning, hochreiter1997long}) that process input tokens sequentially, the transformers can extract relations and importance of individual input tokens in a unit of a whole sentence or a large chunk of words in parallel through attention computation, which offers parallelism for hardware to tap. 

\begin{figure}[t]
\centering
\includegraphics[width=0.45\textwidth]{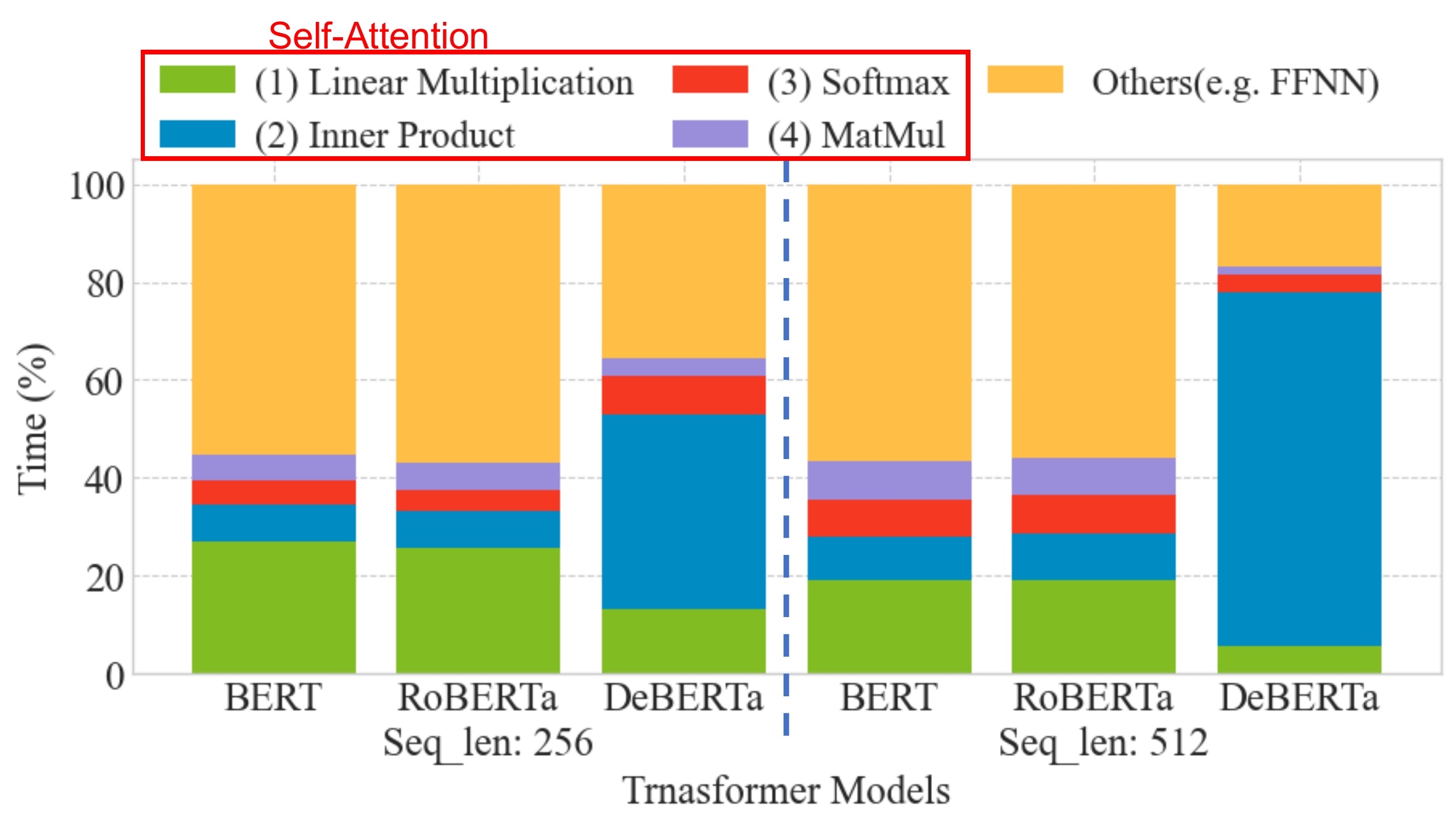}
\vspace{-15pt}
\caption{Inference time breakdown for three  transformers.} 
\label{break_down}
\vspace{-20pt}
\end{figure}

\textbf{Problems.} The highly parallelized self-attention mechanism is compute-intensive, and takes the largest portion of the total inference time of a transformer model. Figure~\ref{break_down} breaks down the inference time of three popular transformer models using two input-sequence lengths (256 and 512). In all cases, the self-attention is the most time-consuming operation, taking 43\%-80\% of total inference time. Also, the portion of the self-attention computation time increases when the input sequence becomes longer.  


To reduce computation of the self-attention mechanism, several recent efforts~\cite{wang2021spatten, ham20203} exclude unimportant input-tokens from the computation. However, these efforts achieve better performance at the cost of losing model accuracy. The accuracy drop is significant, especially in complex tasks that require full-contextual information, such as conversation generation. Some efforts~\cite{elsa, wang2021spatten, lu2021sanger, tambe2021edgebert, ham20203} require a specialized hardware accelerator to achieve the expected performance enhancement.  
Some efforts~\cite{kitaev2020reformer, guskin2021dynamic, kim2022learned} exploit the architectural variances of diverse transformer models and introduce sparsity into the self-attention to reduce computation cost. These efforts leverage the sparsity-based optimizations based upon content-based methods such as locality sensitivity hashing (LSH)~\cite{kitaev2020reformer}, early estimation~\cite{elsa}, cascaded filter~\cite{wang2021spatten}, dynamic sequence length~\cite{guskin2021dynamic}, and learned filter~\cite{kim2022learned}. They show that the time complexity of self-attention can be reduced from 
$\mathcal{O}{(L^2)}$ to $\mathcal{O}{(KL)}$, where $L$ is the input sequence length and \textit{K} is a hyper-parameter as in Top-$K$. However, the practicality of these solutions is questionable, because they require either significant changes in the model architecture or specialized hardware to achieve the expected performance improvement.

In this paper, we propose a more scalable method to reduce attention-computation overhead and  do not require model architecture reconstruction or costly specialized accelerators. Our method is driven by the recent emergence of memory technologies (such as expanded memory via Compute Express Link (CXL)~\cite{cxl} or persistent memory~\cite{Optane:intelweb}) that enable big memory systems at large scales (e.g., the terabyte scale). Such a big memory system can not only accommodate highly memory-consuming applications but also provide opportunities to enable new programming and computation paradigms~\cite{ics21:memoization,asplos20:0sim,amazon_high_mem_inst}. In particular, we trade memory capacity for computation capability to improve the self-attention performance by memoization.

\textbf{Key insight.} Our method is based on the observation that there is similarity in self-attention computation across the inferences of transformer models. Hence, we memoize the results of the most time-consuming tensor computation in the self-attention mechanism  and replace the expensive tensor computation with lightweight searches in a key-value store (KVS). 

Our intuition to use memoization is driven by the similarities commonly found in the input of transformer models (e.g., natural language sentences~\cite{nlp-paper}). For example, \textit{``I like apple.''} and \textit{``I like banana.''} have different meanings, but the important words (that have higher attention scores in the self-attention mechanism) in these two sentences, which are \textit{apple} and \textit{banana}, are obviously in the same position. As a result, the self-attention results (i.e., the attention tensors) between the two sentences can be similar (as a human perceives the structures of the two sentences as similar). Hence, we can reuse the attention tensor calculated by the first sentence for the second.  


\textbf{Challenges.} However, this seemingly intuitive idea faces several challenges to be practically materialized. 
The first challenge lies in finding a proper data representation to extract similarities in self-attention computation. 
Though humans can recognize similarities among sentences, it is unclear if those similarities are reflected numerically in the self-attention computation. To identify similarities from it, we design a proper data representation through embedding. The embedding algorithm must be lightweight: its overhead plus key-value search should be smaller than the cost of self-attention computation, in order to get performance benefits. 


The second challenge is the expensive memory accesses for storing and fetching pre-populated (memoized) tensors in the KVS. To improve the search hit rate and leverage the big memory capacity offered by the big memory system, the KVS should be big and can be pre-populated with attention tensors during the transformer model training. However, as accesses to attention tensors do not have good spatial and temporal locality (e.g., neighbouring attention computations do not have a certain pattern to generate similar tensors), the large key-value search leads to highly sparse memory accesses. 
To make things worse, modern deep learning frameworks like PyTorch require the tensors to be placed in consecutive memory addresses to enable vectorized data accesses for SIMD operations. Therefore, once the tensors are fetched from the pre-populated KVS, the tensors should be copied to a consecutive memory space and then loaded to the processor registers to be used by the attention function. As a result, one tensor fetch generates two memory reads and one write, which may deter 
the performance gain of memoization. 
Hence, it is critical to cut the chain of the cascaded memory accesses for individual tensor searches.


The third challenge is to find the optimal memoization level by considering the tradeoff between inference time (performance) and inference accuracy of transformer model. If performance improvement is the first-priority goal, we can enforce the same amount of self-attention computations to be replaced with memoized attention-tensors at all layers of the transformer model (e.g., all layers replace 50\% of self-attention computations with memoized tensors). But, in that case, some tensors might be replaced with less similar ones and hence the inference accuracy cannot be guaranteed. On the other hand, if there is a strict accuracy requirement, tensors should replace computations only when the expected similarity is high enough. Therefore, the memoization opportunity might be imbalanced across layers in the transformer. Furthermore, the performance penalty in those  layers that have lower chances of successful memoization can be significant because of embedding and search overhead. 
If the search fails to find a matching tensor, the original attention computation should be executed. As a result, the performance can be even worse than when memoization is not used at all. Therefore, there must be a way to determine the memoization effectiveness for individual attentions without actual tensor search to avoid performance loss.

\textbf{Solutions.} To tackle these challenges, we propose an efficient memoization framework (named \textit{\name}).
\name is based on three major innovations: (1) \name uses a lightweight multi-layer perceptron (MLP) model for embedding. Such a model effectively maps input tensors (used as input to self-attention) from a high-dimensional space to a lower one in order to reveal implicit computation similarity. (2) \name eliminates expensive tensor copying through memory mapping between a consecutive virtual-memory space and scattered physical addresses of individual tensors. With this memory mapping technique, we eliminate memory copies completely and achieve over 500$\times$ speedup than tensor fetching in PyTorch. (3) \name employs a performance model that enables selective memoization and avoids unnecessary embedding and search.

In summary, this paper makes the following contributions.
\begin{itemize}[leftmargin=*,noitemsep,topsep=0pt]
    \item We introduce a fundamentally new method to accelerate self-attention for transformer model inference. This method, based on memoization, is motivated by emerging big memory systems and our unique observation that there is implicit similarity in self-attention computation across model inferences.

    \item We identify major challenges to apply big-memory-based memoization to self-attention, and design an end-to-end framework (AttMemo)
    for efficient memoization.

    \item Evaluated with representative transformers (including GPT-2~\cite{gpt}), \name enables 22\% performance improvement on average (up to 68\%) with ignorable loss in inference accuracy. \textcolor{dong}{We also identify rich similarity in self-attention computation across inferences in a large language model with billions of parameters (particularly LLaMA~\cite{touvron2023llama}), demonstrating the potential of using memoization.}
\end{itemize}

\section{Background}

\subsection{Self-Attention Mechanism}\label{self-attention-mechanism}

\begin{figure}[t]
\centering
\includegraphics[width=0.45\textwidth]{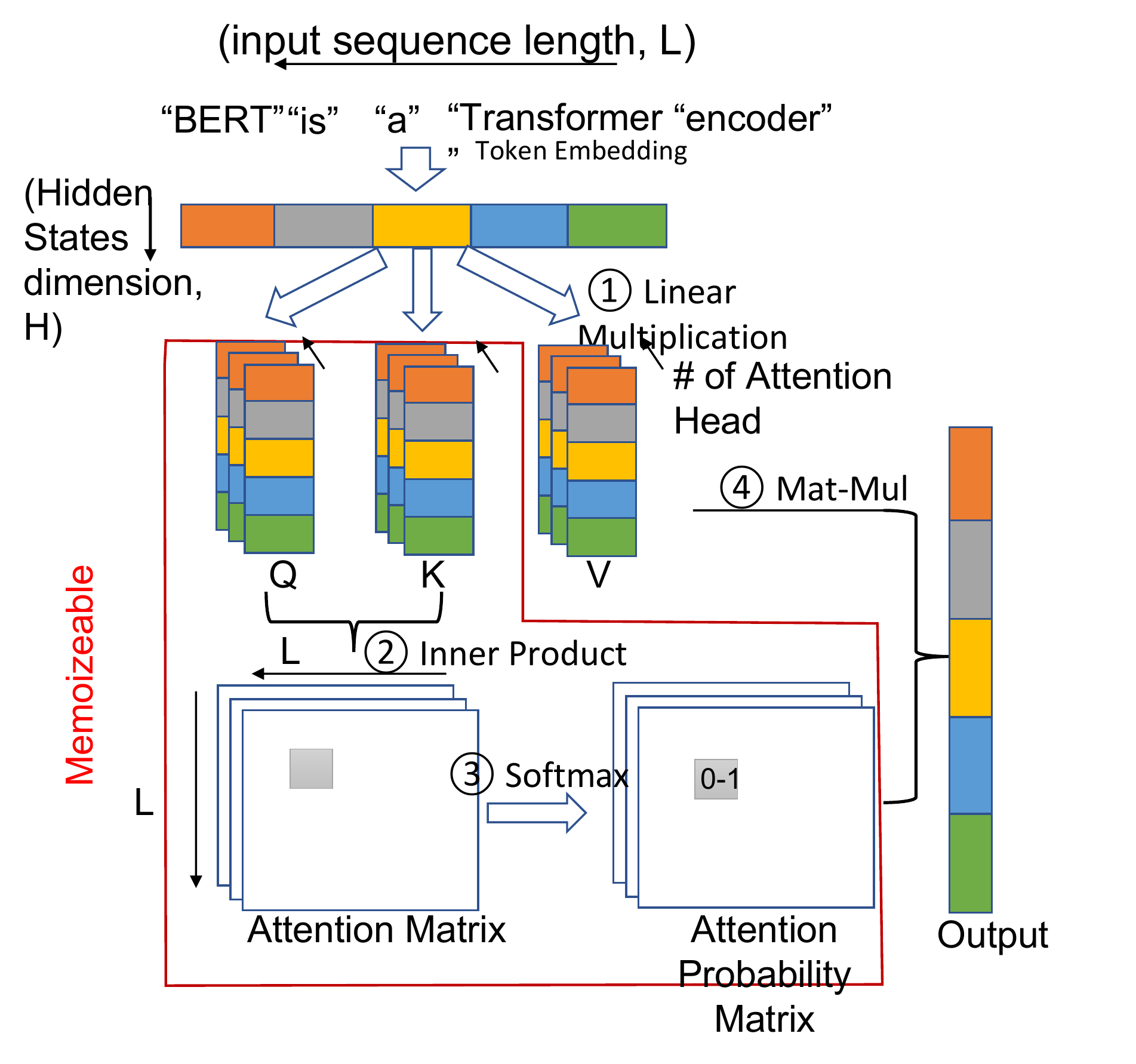}
\vspace{-10pt}
\caption{Self-attention mechanism} \label{attention_workflow}
\vspace{-15pt}
\end{figure}

\textbf{Function description.} The self-attention mechanism is a core building block in a transformer model. Given an input sequence of entities (e.g., words),  self-attention aims to find correlations between different entities of the input in order to indicate the syntactic and contextual structure of the input sequence. Figure~\ref{attention_workflow} depicts the computation involved in a typical self-attention layer. The computation includes four major steps.  In particular, the input sequence of the transformer model is tokenized and embedded into a set of vectors (i.e., the input hidden states). Then, the vectors are multiplied with three weight matrices ($W_Q$, $W_K$ and $W_V$) to generate three intermediate tensors $Q$, $k$, and $V$, shown in \circled{1}. Then, $Q$ and $K$ perform an inner product (shown in \circled{2}), whose result is called the \textit{attention matrix} (AM). The dimensionality of the AM is $L \times L$, where $L$ is the length of the input sequence. Each row of the attention matrix is softmax-normalized, shown in \circled{3}. The normalization result is the \textit{attention probability matrix} (APM). APM is then multiplied with $V$ to get the final output, shown in \circled{4}.  

Figure~\ref{attention_workflow} shows the workflow of a self-attention layer. A self-attention layer is usually followed by post-attention layers, such as feed-forward network and CNNs. A transformer model can have many such combinations stacked on top of each other. For example, the BERT-base model has 12 self-attention layers, and each of them is followed by a feed-forward network and a residential connection. Also, in a self-attention layer, the self-attention mechanism can be split into multiple sub-domains to capture the different aspects of information, which is called multi-head attention.


\textbf{Self-attention similarity.} In essence, self-attention is a mechanism to reveal the relationship between the entities within an input sequence. Each entity is matched with the most relevant other entities from the input, and their relationships are quantified with attention scores through the multiplication of $Q$ and $K$ and normalization by softmax. Since the result of the softmax function is a probability distribution, if the syntactic structure and entity relationship in two input sequences are similar, then it is possible that the probability distributions (i.e., the self-attention results) of the two input sequences are the similar. 

\textbf{Transformer model inference.} While training is normally accelerated with throughput processors such as GPUs, the transformer models have been widely deployed on CPUs for model inference in industry (e.g., Oracle~\cite{oracleCPUtransformer}, HuggingFace~\cite{efficient_inference_on_cpu}, Microsoft~\cite{DBLP:journals/corr/abs-2010-13382} and Intel~\cite{nori2021reduct}). 
Compared with using GPU, using CPU for inference has the benefit of significantly lower production cost (e.g., 16x less shown in Infaas~\cite{cheaper}). Furthermore, for large models (e.g., LLaMA and BLOOM~\cite{Bloom+:GCS62}) that need multiple GPUs or multiple rounds of communications between host CPU and a GPU for each inference~\cite{atc21:zerooffload} because of limited memory per GPU, using CPU (and large CPU memory) does not suffer from the overhead of data movement across GPUs. Our evaluation in Section~\ref{sec:llm_potential} shows that 65-billion-parameter LLaMA using 6 CPU instances on a cloud can perform better than using certain GPU configurations (e.g., 4 GPU instances) by 9\%, \textcolor{dong}{while reducing hardware acquisition cost and cloud cost by 29\% and 80\% respectively.}




\section{Related Work}

\textbf{Applying memoization for performance optimization.} 
Silfa et al.~\cite{silfa2019fuzzy} propose a binary neural network (BNN)-based memoization scheme to adaptively decide when to use memoization on neurons to accelerate RNN training. Xie et al.~\cite{ics21:memoization} build an efficient two-phase LSM tree as a lookup table and leverage the capacity advantage of Intel Optane Persistent Memory to accelerate molecular dynamic simulation. Some studies~\cite{ning2019deep, wu2022drew, ning2019adaptive} empirically prove that  similar computation exists during CNN training and inference, and propose to project similar CNN computation results into buckets by locality-sensitive hashing (LSH) to accelerate the CNN training and inference. 

\textbf{Similarity in DNN computation.} 
Prior works~\cite{ning2019deep, ning2019adaptive,  wu2022drew, silfa2019fuzzy} reveal high computation similarity  in convolution in a single image data or RNN neuron activation in  consecutive time steps. Cao et al.~\cite{deformer} introduce a decoupled transformer architecture dedicated to Question-Answering (QA) tasks. It reuses encoding results in shallow layers for the question part across inferences. 
However, this method is limited to QA where there is explicit duplication of texts which leads to computation similarity.  Srinadh et al.~\cite{google-reuse} reuse attention probability across consecutive self-attention layers for an input text. However, no prior work revealed similarity in attention probability and key intermediate tensors in self-attention layers across input sequences like \name. 

\textbf{Optimizing performance of the transformer.}
Recent studies improve learning-task performance as well as computation efficiency of self-attention mechanisms. Google BERT~\cite{devlin2018bert} is a bidirectional transformer for language modeling. ALBERT~\cite{lan2019albert} 
reduces memory consumption of BERT by sharing weights across self-attention blocks. Adaptive-span~\cite{sukhbaatar2019adaptive} asks each token to perform self-attention with their neighbor tokens within an adaptive window instead of performing full-length self-attention. 
Reformer~\cite{kitaev2020reformer} utilizes random projection-based LSH to project query-key pairs into hash buckets and then computes the attention score of each bucket. 
Wu et al.~\cite{wu2021linear} model attention computation as a maximum inner-product search problem and design a recommendation system based on a single-layer self-attention utilizing the histogram of the encoded items in the input sequence. Ham et al.~\cite{elsa} propose a random projection-based transformer approximation scheme and a domain-specific hardware to accelerate self-attention. Dao et al. \cite{dao2022flashattention} propose an IO-aware tiling mechanism to accelerate forward and backward passes of transformer  training; Kao et al., \cite{flat-asplos23} propose a specialized dataflow-based architecture to accelerate attention computation for long sequences. Different from the above solutions, \name neither change transformer structures nor require specialized hardware. 

\vspace{-5pt}

\section{Motivation}
\label{sec:motivation}


\textbf{Cost of self-attention mechanism.} As explained in Section \ref{self-attention-mechanism}, the self-attention mechanism consists of four key steps. We analyze their computation complexity as follows. The first step ($K$, $Q$, and $V$ projection) requires $3\times L \times H$ multiply-and-accumulate (MAC) computation (where $H$ is the dimension of the token embedding, which is also referred as hidden state). The second step ($Q \cdot V$) requires $L^2\times H$ MAC. The third step (softmax) requires ($L^2$) exponent computation. The fourth step ($V \cdot$ APM) requires $L^2 \times H$ MAC.


We measure the execution time of self-attention mechanism. We use SST2 dataset from GLUE benchmark suite~\cite{glue} and evaluate three transformers (BERT, RoBERTa and ALBERT). The input sequence lengths are 256 and 512. Our evaluation is performed on an Intel Xeon Gold 6252 CPU (24 cores in total). More detailed discussions on the transformer models can be found in Section~\ref{sec:eval_setup}. Figure~\ref{break_down} presents the results. Figure~\ref{break_down} reveals that the self-attention mechanism takes more than 40\% of the total inference time in all the three models. As we increase the input sequence length, self-attention takes a larger portion (up to 83\%). Hence, the self-attention is the performance bottleneck in the transformer inference. 


\begin{figure}[!t]
\includegraphics[width=0.47\textwidth]{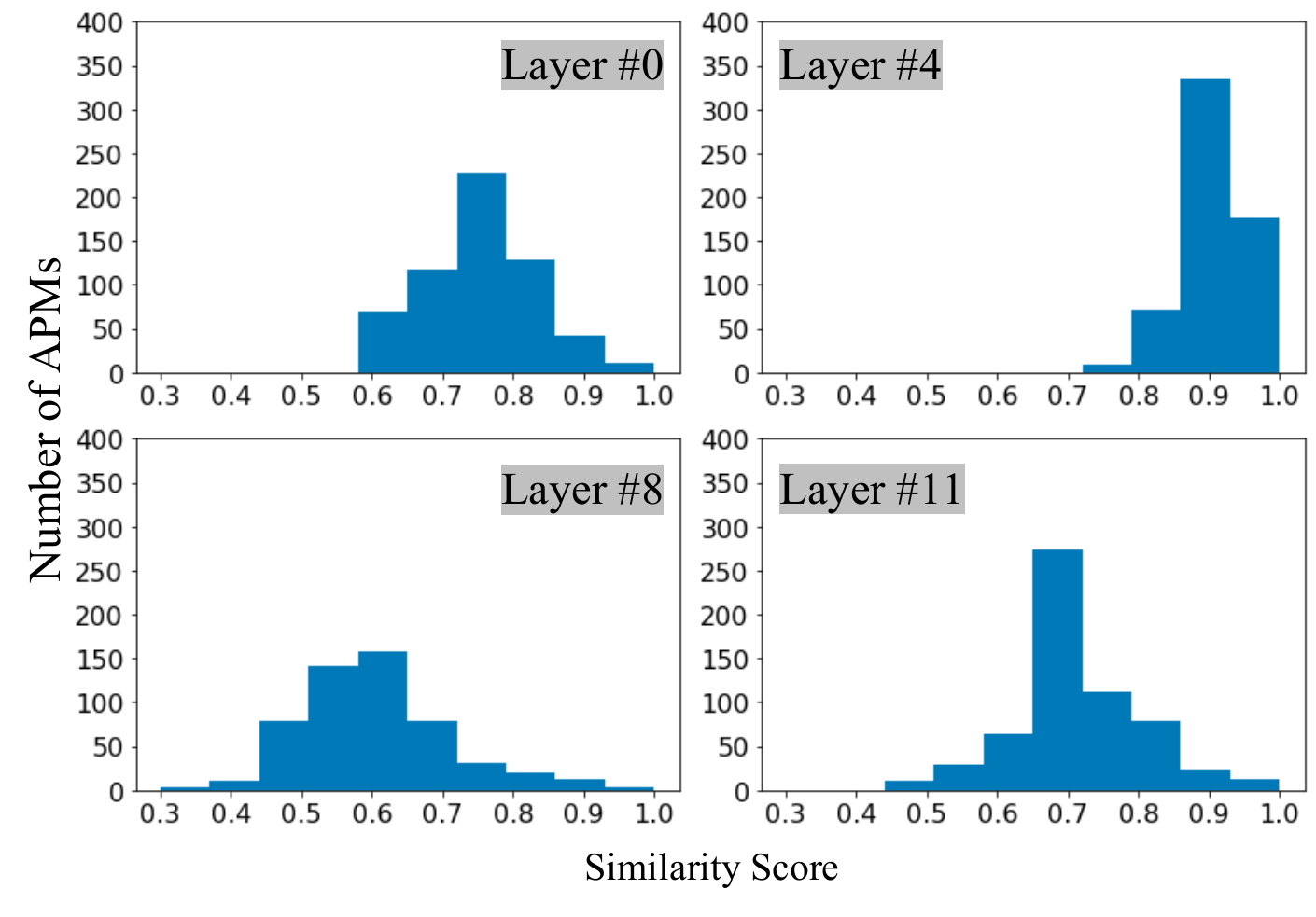}
\vspace{-10pt}
\caption{Distribution of similarity scores in BERT} \label{fig:similarities}
\vspace{-15pt}
\end{figure}  

\textbf{Opportunity in similarities.} 
The success of memoization depends on the existence of computation redundancy, manifested as the similarity between computation results. In the self-attention mechanism, we identify that APMs often show similarity across input sequences. To quantify the similarity, we use a metric based on the total variation (TV) distance~\cite{google-reuse}. The metric (named the \textit{similarity score}) is defined in Equation~\ref{eq:TVD}. 

\begin{equation}
\small
\label{eq:TVD}
\begin{aligned}
SC(A, A') &= 1 - \frac{1}{n} \sum^L_{p=1} TV (A[p, :], A'[p, :]) \\
& = 1 - \frac{1}{n} \sum^L_{p=1} \frac{1}{2} || (A[p, :] - A'[p, :]) ||_1 .
\end{aligned}
\end{equation}

\noindent where $A$ and $A'$ are matrices (APMs); $L$ is the input sequence length; $||.||_1$ denotes $L1$ norm; $A[p,:]$ (or $A'[p,:]$) denotes the $p$th token in the input sequence or the $p$th row in $A$ (or $A'$). Since the attention probability for each token in the input follows a probability distribution, TV falls into [0,1], so as the similarity score. A same definition is used in~\cite{google-reuse}. 


We use BERT~\cite{devlin2018bert} with SST2 from the GLUE benchmark suite~\cite{glue}  as a case study. There are 12 self-attention layers in BERT. We collect APMs from all of them using 60K input sequences in the SST2 training set. Those APMs are used to build an \textit{attention database}. We use another 600 sequences from the SST2 testing set for BERT inferences. During each inference, once we calculate APM for the first attention head of each layer, we search the attention database to find the most similar APM record using the similarity score. 
Figure~\ref{fig:similarities} shows the distribution of similarity scores calculated from those records for four self-attention layers in BERT. 
We have two observations. 

\vspace{2pt}
\begin{itemize}[leftmargin=*,noitemsep,topsep=0pt]
    \item A large percentage of the APMs can find records with a high similarity score (i.e., between 0.7 and 0.9) in the attention database. For example, in Layer 0, 49.2\% of APMs find records with high similarity. 
    
    \item The similarity distribution is different across layers. Hence, we must apply memoization adaptively for high inference accuracy. 
\end{itemize}
\vspace{2pt}

\begin{figure}[!t]
\centering
\includegraphics[width=0.44\textwidth]{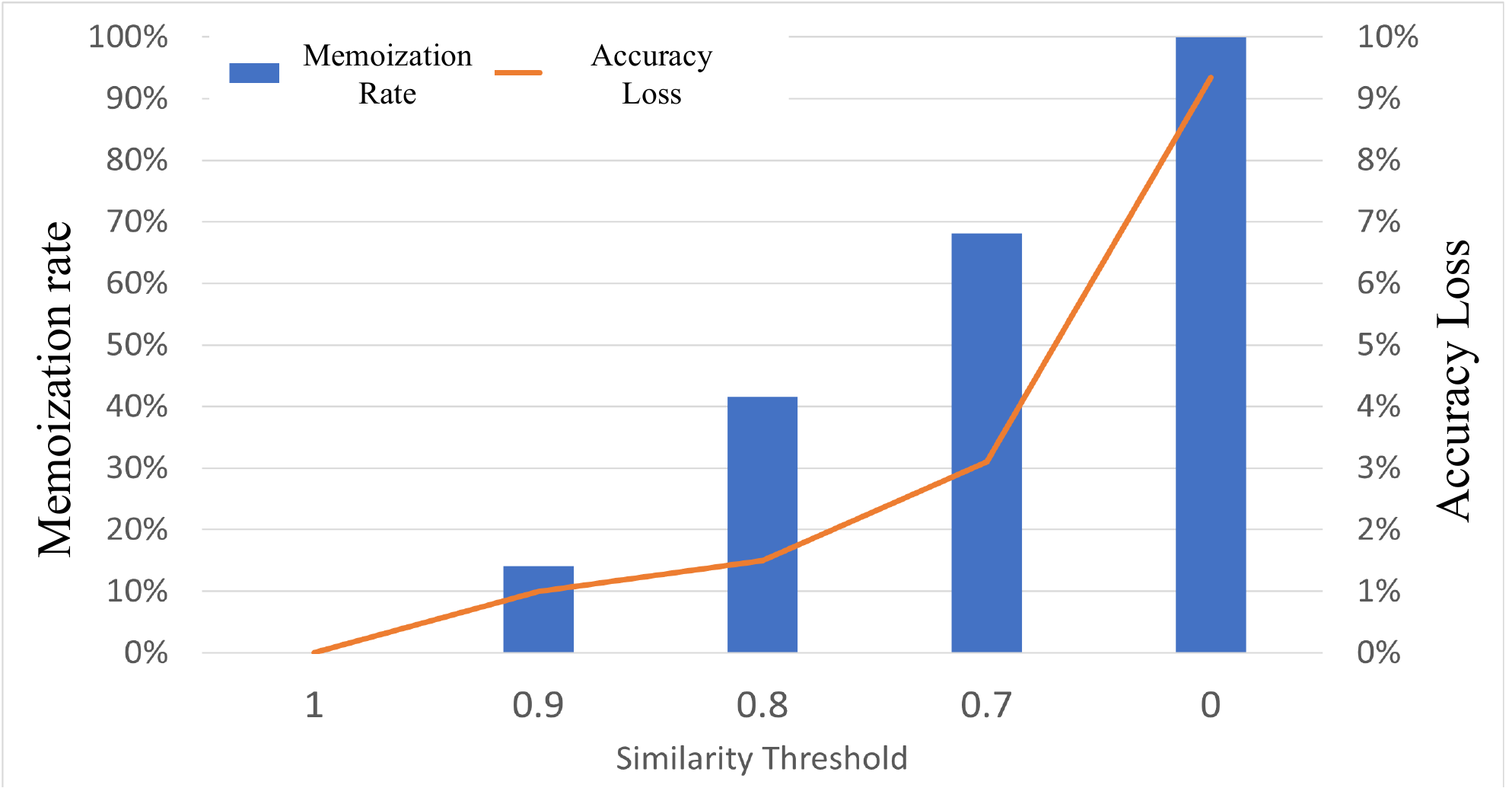}
\vspace{-10pt}
\caption{Impact of memoization on BERT accuracy} \label{acc_sim_drop_exhaustive_search}
\vspace{-10pt}
\end{figure} 


\textbf{Impact of applying memoization on accuracy.} As a preliminary study, we use a similarity threshold to control whether memoization should be used. We use the same attention database and BERT as in Figure~\ref{fig:similarities}. 
We use the search result from the attention database for self-attention only when the similarity score is larger than the threshold. 
We change the similarity threshold from 1 (i.e., no memoization) to 0 (i.e., all memoization), 
and then measure the BERT inference accuracy. In our study, we use a term, \textit{memoization rate}, defined as follows. Assume that there are $N$ input sequences for a
transformer model with $L$ self-attention layers. During the inferences of those $N$ inputs, we count the number of times memoization is \textit{successfully} 
applied to individual layers, denoted as $M$. Then, the memoization
rate ($ms$) is defined as follows.
\begin{equation}
\label{eq:ms}
    ms = M/(N \times L)
\end{equation}

Figure~\ref{acc_sim_drop_exhaustive_search} shows the results. We have two observations.

\begin{itemize}[leftmargin=*,noitemsep,topsep=0pt]
\item As we reduce the threshold, the memoization rate increases (i.e., more self-attention uses memoization). 

\item The accuracy loss can be small even when the memoization rate is high. For example, when the threshold is 0.8 and the memoization rate is 42\%,  the accuracy loss is less than 1.5\%, ignorable in the BERT inference task.

\end{itemize}

\section{System Design}

\subsection{Overview}
\begin{figure}[!t]
\centering
\includegraphics[width=0.48\textwidth]{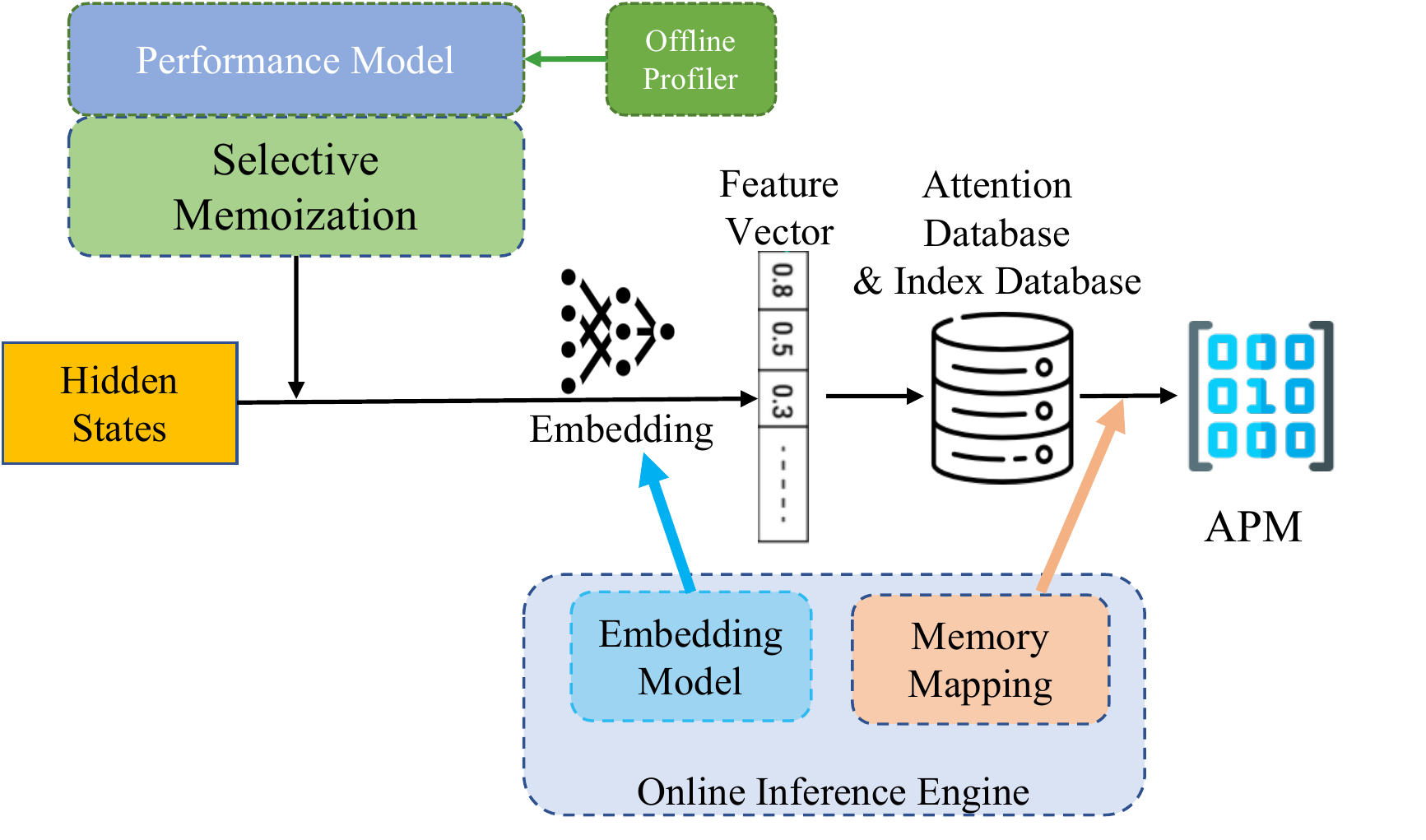}
\vspace{-20pt}
\caption{End-to-end workflow of \name}
\vspace{-10pt}
\label{workflow}
\end{figure}

\name has four major components: attention database, index database, offline profiler, and online inference engine, depicted in Figure~\ref{workflow}. The \textit{attention database} stores pre-computed APMs. \textcolor{dong}{Each APM is associated with a hidden state which is used as the input in the original self-attention to generate the APM. Those hidden states are organized in the \textit{index database} for efficient search (Section~\ref{Construction}). Given a transformer inference request, using a lightweight embedding model (Section~\ref{HiddenStates}), the \textit{online inference engine} embeds the input to each self-attention layer (this input is a hidden state). Then the inference engine uses the embedding result (a feature vector in Figure~\ref{workflow}) as a key to query the index database. Using embedding, \name is able to find semantically similar hidden states in the index database, hence increasing the memoization opportunities. Given the query, the index database returns an index to an APM whose associated hidden state is closest to the query in the embedding space.} That index is used by the inference engine to retrieve the APM from the attention database. After that,  
the inference engine uses a memory mapping technique  (Section~\ref{Construction}) to utilize the returned APM in the remaining inference computation without paying any memory copy overhead. The \textit{offline profiler} is used to build a performance model for selectively applying memoization (Section~\ref{localization}). The selective memoization is necessary, because memoization cannot be always successfully applied but the memoization overhead has to be paid. The offline profiler is based upon the transformer training process to build the performance model to predict whether using memoization at a specific self-attention layer can lead to performance benefit. We discuss details as follows. 
 
\subsection{Hidden State Embedding} 
\label{HiddenStates}

\textcolor{dong}{The input to self-attention is a hidden state, which is the output of the immediately preceding layer of self-attention. Rather than directly using the hidden state as a key to find an APM, \name uses the embedding of the hidden state.}   



\textbf{Why embedding?} 
To match a hidden state with a pre-recorded hidden state to find an APM, we could decide the match based on the calculation of similarity score defined in Equation~\ref{eq:TVD}.  
However, such a match cannot capture semantic similarity between two input hidden-states. Two hidden states may be very different causing a low similarity score but lead to similar APMs. We call these two hidden states, \textit{semantically similar}. The semantically similar hidden-states bring more opportunities for memoization. 

To quantify the semantic similarity, we use an embedding network. When the embeddings of two hidden states are similar (in terms of the similarity score), the two hidden states are matched during a search. The embedding is essentially an internal representation of the input feature (i.e., the hidden state) to \textcolor{dong}{capture semantic similarity}, and the embedding network learns this representation through training such that the hidden states having semantic similarity have similar embeddings. The embedding has been used in the existing work for information retrieval problems~\cite{wei2017selective,song2019polysemous}. Different from the existing work, our objective is to search for hidden states producing similar APMs through embedding. 

Besides identifying hidden states with semantic similarity, embedding allows us to map the hidden state from a higher-dimension space to a lower one. It hence reduces the computation complexity of measuring the similarity. Given an input sequence of a transformer with length $L$ and the dimension of hidden state $H$, the shape of the hidden state tensor is $L \times H$, which is typically in the scale of $O(10^3)$. With appropriate embedding, we can project such a hidden state into a lower dimension (e.g., 128 as the dimension size), reducing computation complexity and search space.

\textbf{Embedding network structure} is critical to the accuracy and efficiency of the search process. We use a Multi-layer Perceptron (MLP) as the embedding model. The MLP is a neural network model with multiple layers of fully-connected nodes. \textcolor{dong}{Our MLP has three layers with tens of thousands of neurons in total and a hidden dimension size of 128.} All neurons are linear neurons ($y = wx + b$). Such an MLP is lightweight. 

Besides MLP, we explore other models for embedding, such as convolutional neural network (CNN) or transformer, which is reported with higher accuracy in some retrieval problems~\cite{koch2015siamese}. However, the CNN and transformer have more computational complexity and require much longer inference time, which easily kills the performance benefit of memoization. For example, with similar accuracy, on Intel Xeon Gold 6252 CPU, the inference time of our MLP for a 64-sequence batch with an input length of 128 takes only 5\,ms while the two-layer CNN- or single-layer transformer-based embedding takes about 100\,ms and 150\,ms respectively.  

\begin{figure}[!t]
\centering
\includegraphics[width=0.47\textwidth]{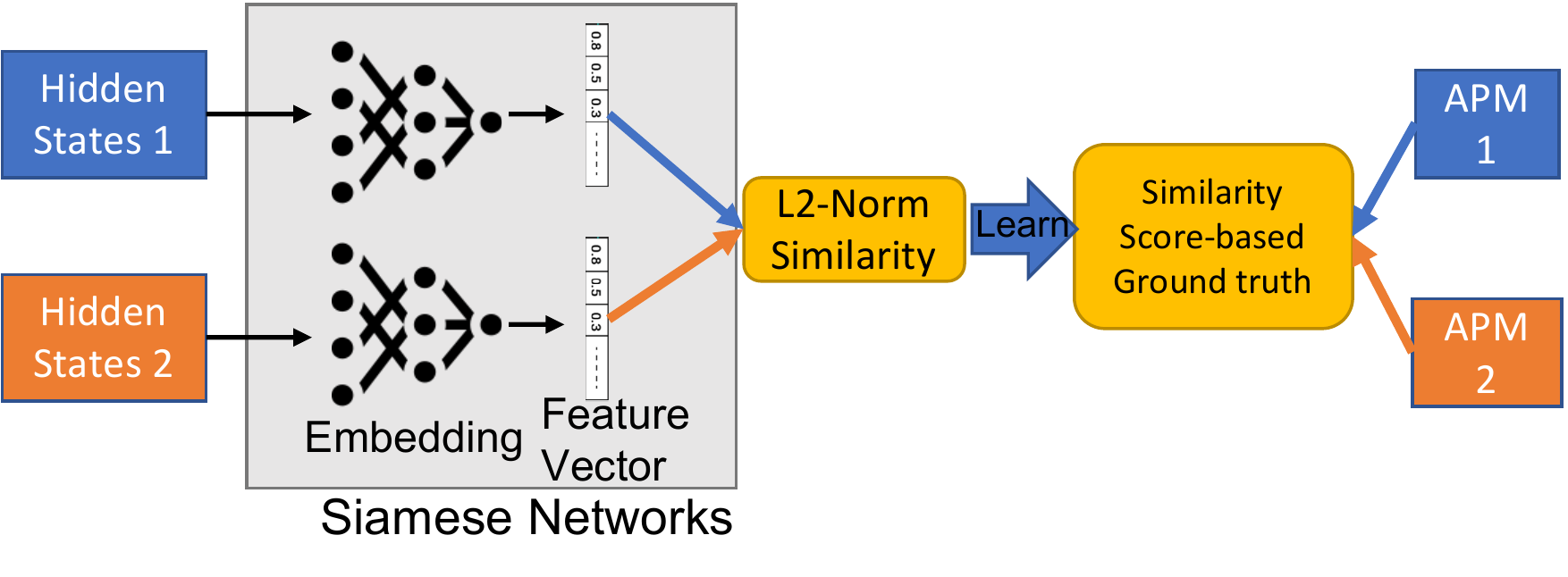}
\vspace{-15pt}
\caption{Siamese network in embedding model training}
\label{Embedding layer training}\vspace{-10pt}
\end{figure}

\textbf{Training} the embedding model is challenging in our case, because of data labeling. Given a big memory system with the potential to accommodate billions of APMs in the attention database,  deciding the similarity between hidden states to label them as similar or not is prohibitively expensive. To address this problem, we use the Siamese network~\cite{koch2015siamese}, a training technique shown in Figure~\ref{Embedding layer training}. The Siamese network is a neural network containing two or more identical sub-networks sharing the same weights and training on the same dataset. In our context, the Siamese network contains two identical embedding models (two MLPs). Once the Siamese network finishes training, one of the embedding models 
is used for memoization. The Siamese network is trained to minimize the distance between hidden states whose embedding results (feature vectors in Figure~\ref{Embedding layer training}) have semantic similarities, while maximizing the distance between hidden states with different semantics.  

During the training, in each iteration, two hidden states are used as input to the two embedding models in the Siamese network. After embedding, the Siamese network calculates the L2-norm between the embedding results from the two embedding models. In addition, the Siamese network measures the similarity score using   
the APMs associated with the two input hidden states. \textcolor{dong}{This similarity score is used as ground truth. Our Siamese network uses the pair-wise 2-norm distance as the loss function to narrow the difference between the L2-norm similarity calculated from the embedding models and the ground truth similarity}. The training iteratively optimizes embedding-model parameters to minimize the loss function. 
With the above training process, we do not need to label hidden states. 



\begin{figure}[t]
\centering
\includegraphics[width=0.43\textwidth]{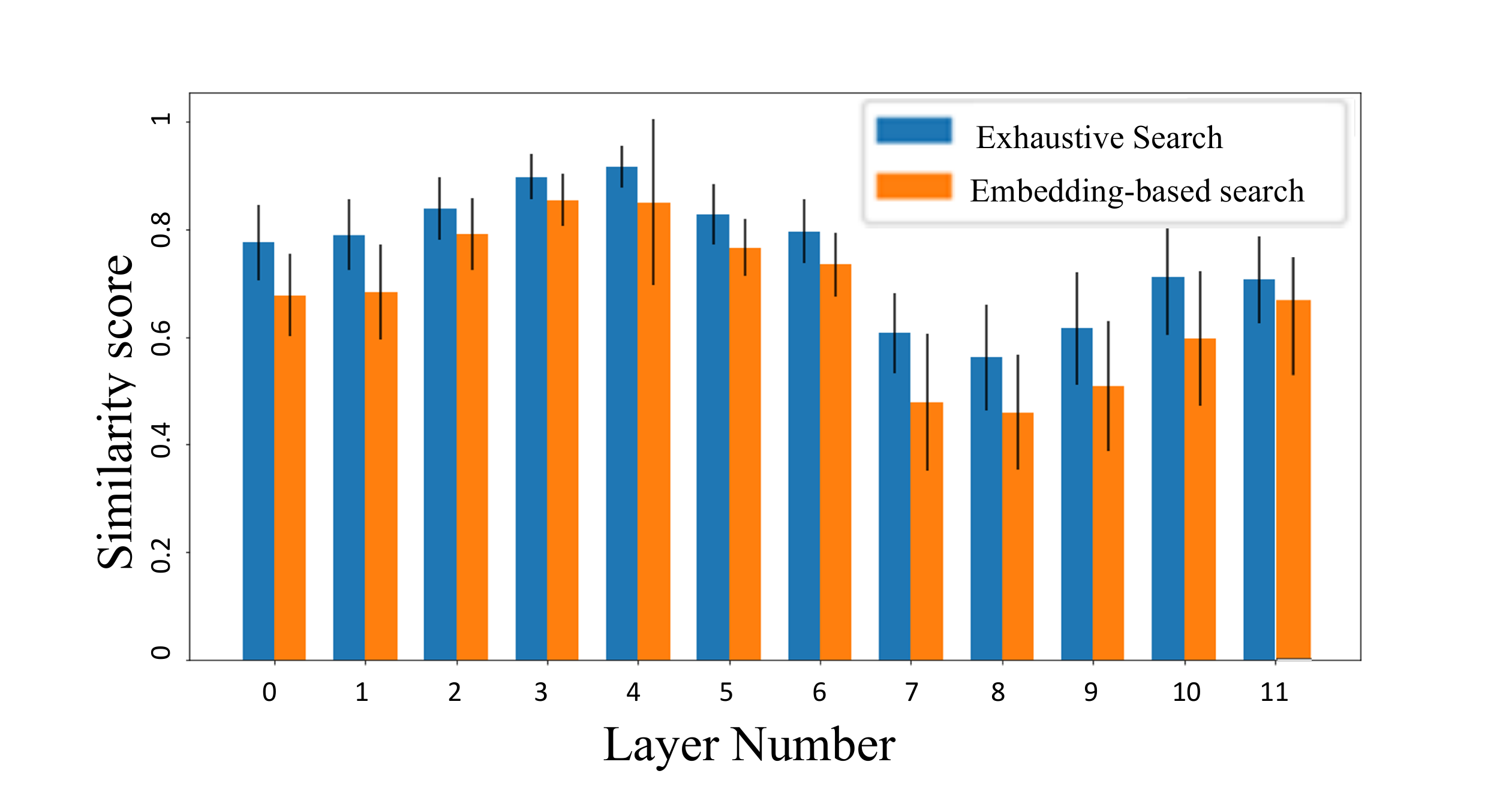}
\vspace{-5pt}
\caption{Exhaustive search vs. embedding-based search.}
\vspace{-10pt}
\label{nn-search-vs-ex-search}
\end{figure}

\textbf{Embedding quality evaluation.} We evaluate the embedding-based search and compare it with an exhaustive search. The exhaustive search provides the best search quality. Given a hidden state as input, the exhaustive search returns a hidden state whose corresponding APM is the most similar to the APM associated with the input hidden state. We use a 12-layer BERT model, build the attention database using 60,000 input sequences from SST-2 dataset from the GLUE Benchmark suite, and use 640 input sequences for the evaluation. 
The embedding-based search is effective, shown in Figure~\ref{nn-search-vs-ex-search}. The average difference in similarity score between the exhaustive search and embedding-based NN search is only less than 0.1. Furthermore, the exhaustive search takes 1.5\,s on average for each search, while the embedding-based search takes only 5 ms, which leads to 300$\times$ speedup. 


\subsection{APM Construction} \label{Construction}

\textcolor{dong}{To avoid expensive search of APMs, we build an index database where indices to pre-computed APMs are searched and organized according to the semantic similarity of  hidden states. The indices in the index database are organized as a hierarchical tree structure to accelerate the search.} We use Faiss~\cite{johnson2019billion}, a library for efficient similarity search using Hierarchical Navigable Small Worlds~\cite{DBLP:journals/corr/MalkovY16} (HNSW, a nearest-neighbour search algorithm) in the index database. With Faiss, searching 100K vectors with  vector-dimension size of 128 takes less than 0.5\,ms, which is 360$\times$ and 10$\times$ shorter than the time for self-attention and embedding respectively. Hence, search does not create a performance bottleneck for memoization.

\begin{figure}[t!]
\begin{center}
% ...
\includegraphics[width=0.40\textwidth]{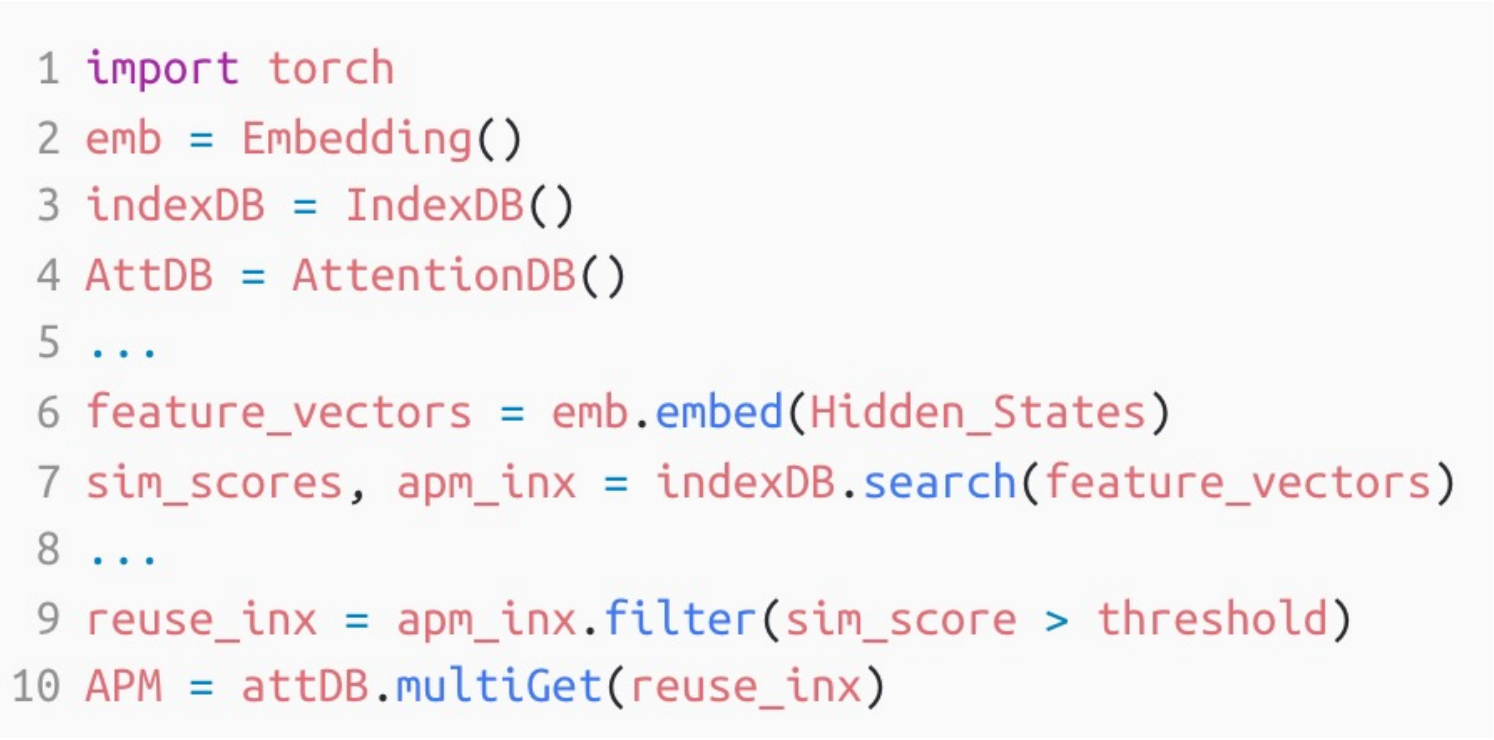}
\end{center}
\vspace{-1.2em}
\caption{Code for APM retrieval from the attention database}
\label{fig:mem_copy_problem}
\vspace{-20pt}
\end{figure}

Figure~\ref{fig:mem_copy_problem} shows the workflow of retrieving APMs from the attention database. A batch of hidden states is embedded (Line 6). The embedding results (\texttt{feature\_vectors}) are used to query the index database (Line 7). \textcolor{dong}{Once the query finishes, the indices \textit{closest} to the \texttt{feature\_vectors} in the tree structure of the index database are returned. The distance between a returned index and a \texttt{feature\_vector} (as an input to the index database) is measured using the Euclidean distance by HNSW. When the distance is larger than a threshold determined by HNSW (Line 9), the corresponding index is returned and used to retrieve an APM from the attention database. APMs are retrieved as a batch (Line 10) to be used in the remaining self-attention.} 


\textbf{Performance problem due to memory copy.} The above process introduces a performance problem due to memory copy. In particular, in the attention database, the APMs are stored in a 
large memory space. During the APM retrieval, APMs are sliced from this space, and then gathered into another contiguous memory space as a tensor for the following computation in self-attention. The APM gathering is necessary in order to exploit SIMD parallelism for high performance. However, APM-gathering copies scattered APMs from the attention database into a contiguous memory space, which introduces nontrivial overhead. For example, using BERT, when the input sequence length is 512, gathering 64 scattered APMs as a batch for a self-attention layer takes 731\,ms on an Intel Optane-based big-memory platform (see Section~\ref{sec:eval_setup} for hardware details), which is 1.45$\times$ higher than the performance cost of self-attention itself. Hence, we must address this performance problem in order to enable the performance benefit of memoization. To address this problem, one could change the machine learning (ML) framework (such as PyTorch) to work on tensors with nonconsecutive memory layouts. However, this requires bookkeeping of tensor allocation and careful manipulation of tensor pointers, which is difficult. 

\begin{figure}
\centering
  \subfloat[][Initial State ]{\includegraphics[width=0.25\textwidth]{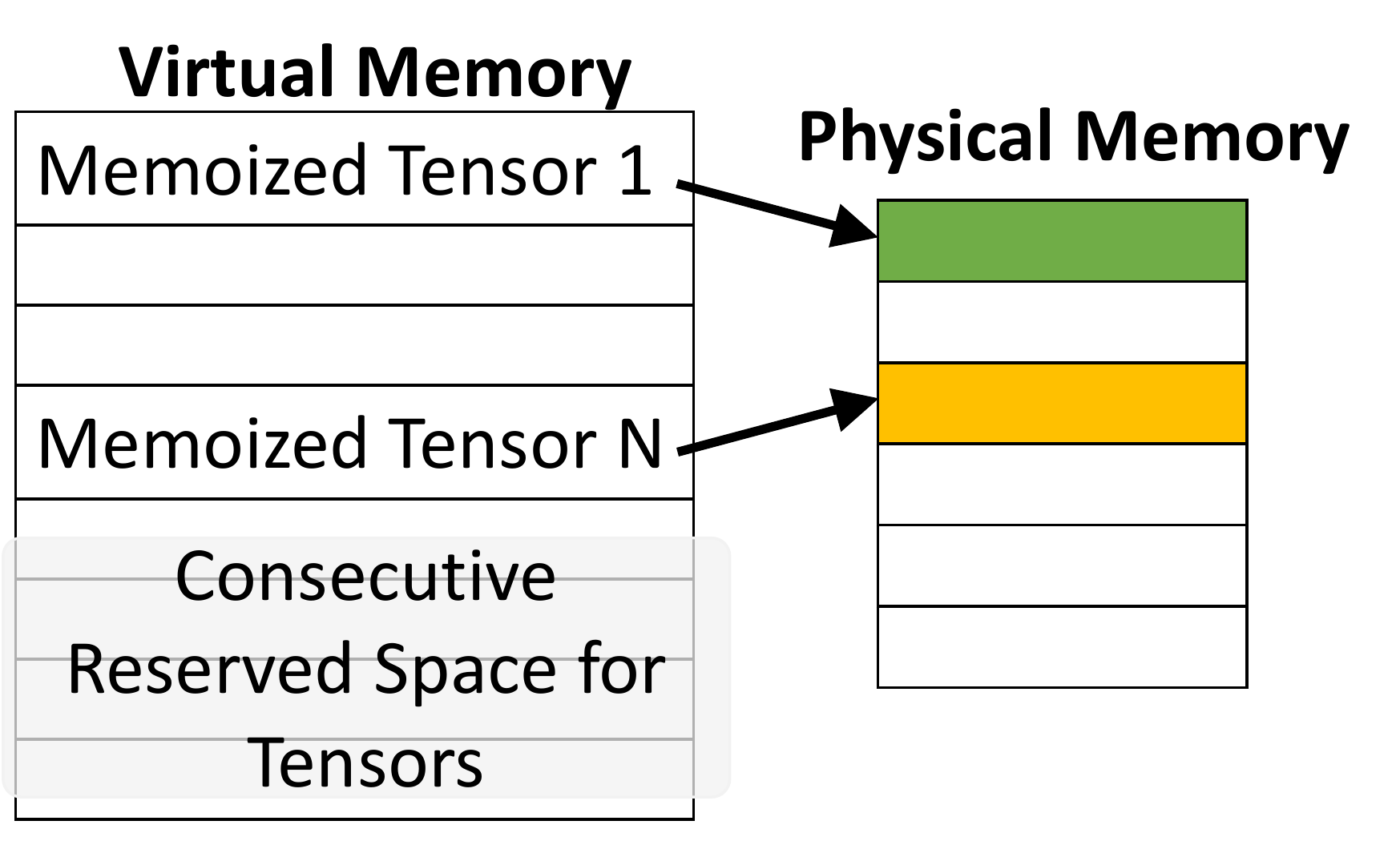}}\\\vspace{-10pt}
  \subfloat[][Copy-based]{\includegraphics[width=0.23\textwidth]{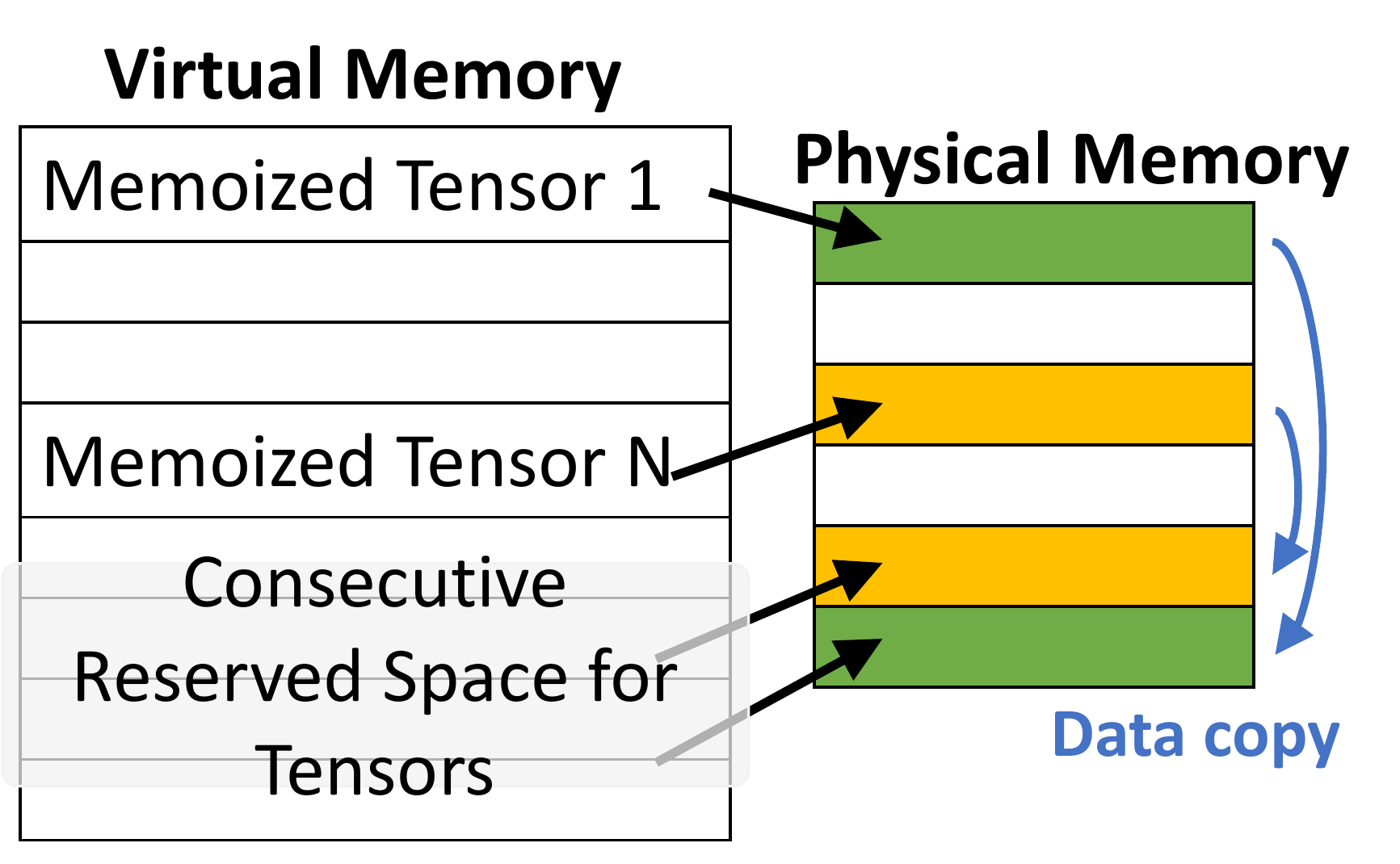}}
  \subfloat[][Mapping-based]{\includegraphics[width=0.25\textwidth]{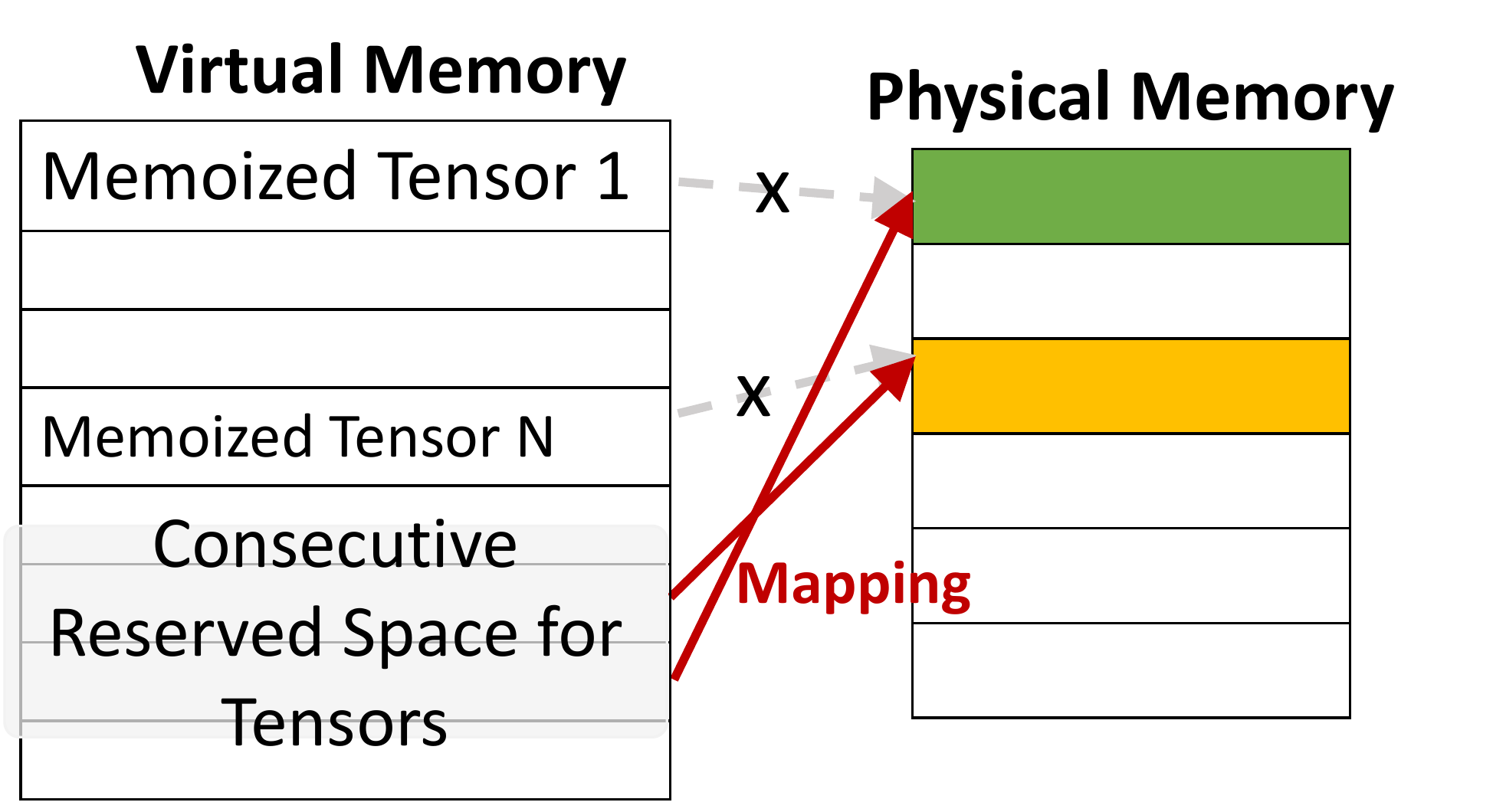}} 
\caption{APM gathering: copy-based vs. mapping-based} \label{fig:collection}
\vspace{-20pt}
\end{figure}


\textbf{Solution based on memory mapping.} We introduce a solution based on memory mapping. When constructing the attention database, each APM is stored as a file object in memory. When retrieving APMs as a batch, APM file objects are mapped into a contiguous virtual-memory space as a tensor without causing memory copy. After self-attention, the file objects are unmapped. Figure~\ref{fig:collection} depicts the difference between the traditional memory copy- and our memory mapping-based approaches. In this figure,  the pages mapped to two APMs (green and yellow boxes) are gathered into  consecutive memory.


\textbf{Performance analysis.} Our solution replaces the expensive memory copy with page table mapping. Behind the scene, the operating system inserts or updates page table entries (PTE) such that the consecutive virtual addresses can be translated to individual APMs' physical addresses. As different APMs are used at different layers of the transformer, the mapping is updated at each layer. However, the PTEs can be reused 
across layers. Once the initial mapping happens in a prior layer, the same PTEs mapped to the consecutive virtual addresses  
can be simply updated with the physical addresses of APMs of the following layers without additional PTE insertion or removal.

The page table update does not incur extra overhead,  because the APMs found from the attention database have to be mapped to the page table even when the ML framework mandates memory copy. Also, as each APM is typically several tens of MBs, this tensor-wise memory mapping does not cause page fragmentation. TLB might be thrashed. However, as APM accesses do not have a high locality (see Section~\ref{eval:tensor_db}), we cannot expect a benefit from TLB in memoization. With the dynamic memory mapping, we can remove memory copy completely and our experimental results show more than 500$\times$ speedup over unoptimized APMs fetching in PyTorch.

\subsection{Selective Memoization} \label{localization}

Applying memoization successfully to self-attention (i.e., finding a similar APM in the attention database and mapping APM to self-attention) leads to performance benefits. However, if a similar APM cannot be found and the memoization cannot be successfully applied, then there is no performance benefit and the embedding and search overhead cannot be covered, which causes performance loss. Among all self-attention layers in a transformer, there should be enough opportunities to successfully apply memoization, such that the overhead can be offset by the performance benefit. If there is no enough opportunity, then the original self-attention computation should happen instead of trying memoization with embedding and searching, such that there is no performance loss. 

Our observation in Section~\ref{sec:motivation} (Figure~\ref{fig:similarities}) reveals that the memoization opportunity is different from one self-attention layer to another. We study how to use performance modeling to guide the application of memoization, such that there is no performance loss (if there is no performance benefit). We apply the memoization at the granularity of a self-attention layer: a layer either applies memoization to all of its attention heads or uses no memoization at all. We do not choose a finer granularity (e.g., an individual attention head) to apply memoization, because of the following reason: the attention heads in the same layer are reported to have high modeling redundancy \cite{google-reuse, bian-etal-2021-attention, DBLP:journals/corr/abs-1905-09418} and tend to make the similar decision on using memoization; embedding and search for multiple attention heads can lead to larger performance overhead. 

We use performance modeling to apply memoization selectively.



\textbf{Performance model} quantifies the performance benefit of memoization, excluding the overhead (including embedding and searching). Given a layer $i$ in a transformer and a batch of input sequences (the batch size is $N$), the performance benefits $PB_i$ after applying memoization is formulated in Equation~\ref{constrains}.
\vspace{-10pt}

\begin{equation}
\label{constrains}
    PB^i = T^i_{Atn} \times \alpha^i - T^i_{overhead}
\end{equation}

\noindent where $T^i_{Atn}$ and $T^i_{overhead}$ are the execution time of the self-attention without memoization in the layer $i$, and the overhead of memoization, respectively, for  $N$ input sequences. $\alpha^i$ is the memoization rate (calculated with $L=1$ in Equation~\ref{eq:ms}). We want $PB^i > 0$.

\textbf{How to build the performance model.} The performance model is constructed during the training of the transformer model. Given a training dataset with $N$ input sequences, for each input sequence, memoization is applied to each self-attention layer. Then, we check whether the memoization brings performance benefit in each layer to decide $\alpha^i$ for each layer. 

\textbf{How to use the performance model.} We use $\alpha$ during online transformer-model inference to guide the application of memoization. The training dataset and online inference dataset of the transformer should have similar properties in order to ensure that the transformer model has meaningful high inference accuracy. Hence, $\alpha$ measured during the training process can be used online to guide memoization. 

During the online inference, a batch of inference requests (or a batch of sequences) is fed into the transformer model. Before inferences happen, we estimate $T_{Atn}$ and $T_{overhead}$ based on the length of those input sequences. The estimation is based on the approximate linear scaling of $T_{Atn}$ and $T_{overhead}$ measured with the training dataset. The scaling factor is the ratio of the total length of inference sequences to the total length of training sequences. 
Given $T_{Atn}$, $T_{overhead}$, and $\alpha$, we use Equation~\ref{constrains} to calculate the performance benefit ($PB$) for each layer. For a layer $i$, only when $PB^i > 0$, we apply memoization.

\textbf{Impact of memoization on inference accuracy.} The performance model does not consider the impact of memoization on transformer inference accuracy. \name considers that by introducing a threshold (named \textit{memoization threshold}; see Line 9 in Figure~\ref{fig:mem_copy_problem}) to ensure the same inference accuracy as the inference without memoization. Only when the similarity score (returned during the search process in the index database) is larger than the threshold, the memoization will be applied.
When the threshold-based control and performance-based control have divergent decisions on whether memoization should be used, we do not use memoization. We expect that the user specifies the threshold as a model hyperparameter to guard inference accuracy, similar to the case that the user specifies other hyperparameters for model inference (\textcolor{dong}{e.g., the number of quantization level for quantization-based model inference}). But an autotuner~\cite{9139814, 10.1145/3453483.3454109, 10.1145/3437801.3441621} can be employed to automatically decide an appropriate threshold.

\section{Evaluation}

\subsection{Experimental Setup}

\textbf{Hardware Platform.} We evaluate \name on a server equipped with two Intel Xeon Gold 6252N 24-core processors running Linux 5.15.0. Each socket has 12 DIMM slots, six of which are for 16 GB DDR4 DRAM (96 GB in total) and another six are for 128 GB Intel Optane DC modules (768 GB in total). The platform provides up to 1.6 TB of heterogeneous DRAM/Optane memory. We use the memory mode of Optane for heterogeneous memory management (i.e., using DRAM as a hardware-managed cache for Optane). 

\noindent\textbf{Software Platform.} We implement \name on top of PyTorch 1.11 with oneDNN (previously known as MKLDNN) support. In the evaluation, all 48 CPU cores are used. We use the computation-based transformer inference (without memoization) using all 48 cores and oneDNN as the baseline.

\noindent\textbf{Evaluation models.} We evaluate \name with transformers listed in Table~\ref{model_size}. 
The input datasets we used in the evaluation are SST-2 dataset from the GLUE~\cite{glue} benchmark suite for the encoder-based and WikiText-v2~\cite{merity2016pointer} for the decoder-based transformers.

\label{sec:eval_setup}
\begin{table}[t!]
\caption{Transformer models in our evaluation}
\vspace{-10pt}
\label{model_size}
\scalebox{0.9}{
\small
\begin{tabular}{lllll}
\hline
Models  & BERT~\cite{devlin2018bert}  & RoBERTa~\cite{liu2019roberta} & DeBERTa~\cite{he2020deberta} & GPT-2~\cite{gpt} \\ \hline
\# Parameters & 110 M & 123 M   & 100 M   & 110 M \\ \hline
\end{tabular}
}
\vspace{-10pt}
\end{table}

\vspace{-10pt}
\begin{table}[t!]
\caption{Memoization threshold settings}
\label{threshold_setting}
\vspace{-8pt}
\scalebox{0.85}{
\begin{tabular}{lrrrr}
\hline
Model         & \multicolumn{1}{l}{BERT} & \multicolumn{1}{l}{RoBERTa} & \multicolumn{1}{l}{DeBERTa} & \multicolumn{1}{l}{GPT-2} \\ \hline
Conservative & 0.98                     & 0.97                        & 0.9995                      & 0.99950                   \\
Moderate      & 0.97                     & 0.96                        & 0.9993                      & 0.99935                   \\
Aggressive    & 0.96                     & 0.95                        & 0.9990                      & 0.99920                   \\ \hline
\end{tabular}
}
\vspace{-10pt}
\end{table}

\begin{table}[t!]
\centering
\caption{Pre-populated attention database size, embedding training time, and index-database building time.}
\vspace{-10pt}
\label{evaluation_detail}
\scalebox{0.8}{
\begin{tabular}{l|lll|lll}
\hline
Input Sequence Length            & \multicolumn{3}{c|}{512}      & \multicolumn{3}{c}{1024}        \\ \hline
\# of Sequences            & 4K      & 6K        & 8K      & 1K        & 1.5K      & 2K      \\
Pre-populated DB Size (GB) & 575     & 855       & 1130    & 630       & 940       & 1250    \\
Embed. Training Time (h)   & $\sim$1 & $\sim$1.5 & $\sim$3 & $\sim$1.2 & $\sim$1.4 & $\sim$2 \\
Indexing Time (s)          & 192     & 278       & 454     & 128       & 176       & 384     \\ \hline
\end{tabular}
}
\vspace{-15pt}
\end{table}






\begin{figure*}[!t]
    \centering
    \includegraphics[width=0.99\textwidth]{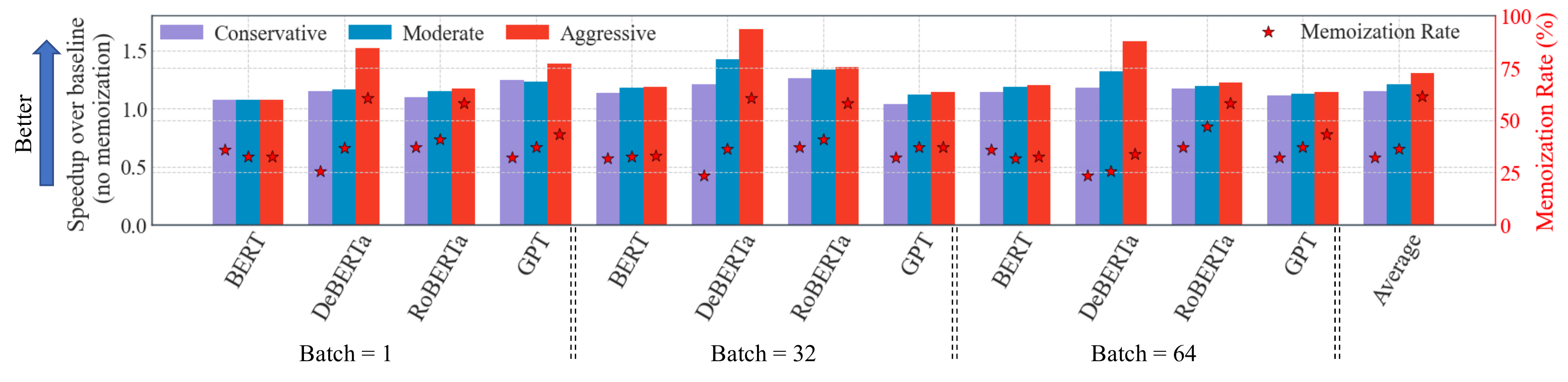}
    \vspace{-15pt}
    \caption{Inference speedup with different batch sizes and memoization levels. For BERT, DeBERTa and RoBERT, the input sequence length is 512; for GPT, the input sequence length is 1K. }
    \label{overall:sub1}
\end{figure*}


    



\subsection{Overall Performance}\label{overall-performance}

We evaluate the performance with two 
typical input sequence lengths: 512 for BERT, RoBERTa and DeBERTa, and 1024 for GPT-2. For each input length, we evaluate three typical batch sizes: 1, 32 and 64. For each batch size, we evaluate three levels of memoization: conservative, moderate and aggressive with respect to the memoization threshold, as depicted in Table~\ref{threshold_setting}. We collect APMs to build and index the attention database using 8K, 6K, or 4K sequences for BERT, RoBERTa, and DeBERTa  and using 2K, 1.5K, and 1K sequences for GPT-2 (see Table~\ref{evaluation_detail}).  

Figure~\ref{overall:sub1} presents the end-to-end performance improvement over the baseline where no memoization is applied. In general, \name brings 22\% speedup on average (up to 68\%). The figure reveals that the performance benefit of \name is more significant when the batch size increases from 1 to 32: the average speadup increases from 19.57\% to 25.71\%. The reason is that the memoization opportunity is higher in the larger batch size, and hence more computations can be replaced with database lookup. When the batch size increases to 64, the performance benefit slightly decreases from 25.71\% to 21.43\%. This is because the embedding model takes more time to embed the hidden states. As a result, the performance benefit of memoization is slightly diminished.

Among the tested models, DeBERTa shows the largest speedup. This is because DeBERTa uses modified (optimized) self-attention layers for better accuracy which are more computationally expensive and take a larger portion of overall inference time. Thus, memoization improves performance better. This sheds the light on the optimization of a transformer model for both accuracy and performance by applying our memoization scheme. 

\textbf{Performance breakdown.}
We break down a single  memoization-based self-attention time in BERT into embedding time, search time, memory mapping time, and other computation. The batch size and input sequence length are 64 and 512 respectively. Table~\ref{tab:perf_breakdown} shows the results. The embedding is the most time-consuming, which highlights the importance of using a lightweight embedding model.

\begin{table}[!t]
\caption{Performance breakdown for a self-attention in BERT}
\label{tab:perf_breakdown}
\vspace{-10pt}
\small
\begin{tabular}{ccc}
\hline
Tims (ms)                                                                                          & \begin{tabular}[c]{@{}c@{}}With \\ Memoization\end{tabular} & \begin{tabular}[c]{@{}c@{}}Without \\ Memoization\end{tabular} \\ \hline
Embedding                                                                                          & 38.4                                                        & N/A                                                            \\
Searching                                                                                          & 1.0                                                         & N/A                                                            \\
APM Fetch                                                                                          & 15.1                                                        & N/A                                                            \\
Q computation(1)                                                                                   & N/A                                                         & 32.2                                                           \\
K Computation(1)                                                                                   & N/A                                                         & 32.2                                                           \\
V Computation(1)                                                                                   & 32.2                                                        & 32.2                                                           \\
APM Computation(2) (3)                                                                             & N/A                                                         & 423.3                                                          \\
\begin{tabular}[c]{@{}c@{}}Other Computation\\ (e.g.  APM $\cdot$ V (4) )\end{tabular} & 163.6                                                       & 47.3                                                           \\
Total                                                                                              & 250.3                                                       & 567.2                                                          \\ \hline
\end{tabular}
\vspace{-5pt}
\end{table}

\subsection{Memoization Quality}

\begin{table}[!t]
\caption{Inference accuracy before and after applying memoization. The batch size is 32.}
\vspace{-10pt}
\scalebox{0.78}{
\begin{tabular}{cccccc}
\hline
\%            & Baseline         & \begin{tabular}[c]{@{}c@{}}AttMemo-\\ \textbf{Conservative}\end{tabular} & \begin{tabular}[c]{@{}c@{}}AttMemo-\\ \textbf{Moderate}\end{tabular} & \begin{tabular}[c]{@{}c@{}}AttMemo-\\ \textbf{Aggressive}\end{tabular} & \begin{tabular}[c]{@{}c@{}}Average\\ Diff.\end{tabular} \\ \hline
BERT          & 93.1             & 92.0                                                            & 92.0                                                        & 91.8                                                          & -1.1                                                    \\
RoBERTa       & 94.8             & 93.2                                                            & 92.6                                                        & 90.4                                                          & -2.7                                                    \\
DeBERTa       & 95.0             & \textcolor{red}{95.5}                                                            & \textcolor{red}{95.2}                                                       & 90.5                                                          & -1.2                                                    \\
Average Diff. & N/A & -0.7                                                           & -1.0                                                       & -3.3                                                         & -1.67                                                   \\ \hline
\end{tabular}
}
\vspace{-15pt}
\label{memoziation_quality}
\end{table}

We measure \namenospace's memoization quality through the inference accuracy. Table~\ref{memoziation_quality} shows the accuracy loss compared to the baseline (no memoization). While the accuracy is maintained consistently across all aggressiveness levels, more aggressive memoization shows a slightly higher accuracy loss (3\%) compared to the conservative and moderate ones (about 1\%). 
There is a clear trade-off between accuracy and performance. 
However, in some models like DeBERTa, we find that a noticeably high memoization-rate can be achieved along with improved accuracy (see the red numbers in Table~\ref{memoziation_quality}) when conservative or moderate-level memoization is used. \textcolor{dong}{Such accuracy improvement may come from the removal of computation redundancy and its interaction with self-attention.}






\subsection{APM Reuse Analysis in Attention Database}\label{eval:tensor_db}
We study how often each APM in the attention database is reused across input sequences. Getting such information is important to understand the importance of using a big memory system to accelerate self-attention with memoization. 

\begin{figure}[!t]
\includegraphics[width=0.47\textwidth]{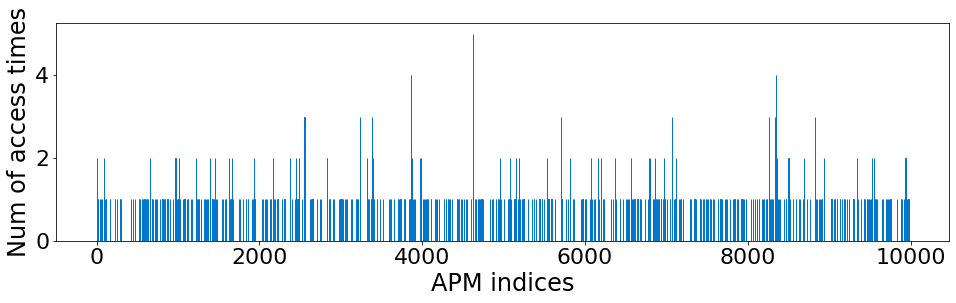}
\vspace{-15pt}
\caption{APM reuse analysis in an attention database.} \label{reuse_freq}
\vspace{-10pt}
\end{figure}

We build the attention database by storing APMs from the first self-attention layer of BERT during the BERT training using 10,000 sequences from the SST-2 dataset. For inferences, we use 640 sequences from the testing set in the same dataset. Figure~\ref{reuse_freq} shows the access times of individual records (APMs) in the attention database. The figure reveals that most records are re-used only once or twice. All of them are re-used no more than six times. There is no ``hot'' record. The traditional memoization work \cite{md-pm2020, ning2019deep,silfa2019fuzzy} heavily relies on a high reuse-rate to be successful, and does not need a big memory system. However, to apply memoization to self-attention, there is no such high APM reuse and we must use a big attention-database to enable success.




\subsection{Efficiency of Mapping-based APM Gathering}

\begin{table}[!t]
\caption{Comparison of APM fetch latency between the memory copy- and mapping-based approaches}
\vspace{-10pt}
\scalebox{0.8}{
\begin{tabular}{c|cccc}
\hline
\begin{tabular}[c]{@{}c@{}}Sequence\\ Length\end{tabular} & \begin{tabular}[c]{@{}c@{}}Batch \\ Size\end{tabular} & \begin{tabular}[c]{@{}c@{}}Mem  Copy\\ (s)\end{tabular} & \begin{tabular}[c]{@{}c@{}}Mapping\\ \& Unmapping (s)\end{tabular} & Speedup \\ \hline
\multirow{3}{*}{256}                                      & 1                                                     & 52                                                          & 0.12                                                               & 433.3x    \\
                                                          & 32                                                    & 104                                                         & 0.l9                                                               & 547.3x    \\
                                                          & 64                                                    & 106                                                         & 0.33                                                               & 321.2x    \\ \hline
\multirow{3}{*}{512}                                      & 1                                                     & 111                                                         & 0.25                                                               & 444.0x    \\
                                                          & 32                                                    & 548                                                         & 0.19                                                               & 2884.2x   \\
                                                          & 64                                                    & 731                                                         & 0.33                                                               & 2215.1x  
\\ \hline
\end{tabular}
}
\vspace{-10pt}
\label{mapping_speedup}
\end{table}

We evaluate the efficiency of the mapping-based APM gathering. We compare it with the copy-based approach.
Table~\ref{mapping_speedup} shows the results. In general, the memory mapping is at least 321$\times$ faster the memory copy.
The superior performance of the mapping-based approach is sourced from the completely removal of data copy. 


\subsection{Impact of Selective Memoization}
We evaluate the impact of selective memoization
on inference time and memoization rate. Table~\ref{selective_memoization} shows the results. The input sequence length is 512. 
With selective memoization, \name improves the performance of all models by 3.0\%-12.3\%. The results show that even when we give up some memoization (on purpose), we can have better performance because of the removal of unnecessary embedding and search overhead. 
Interestingly, for GPT-2, even if we apply selective memoization, the memoization rate increases. This is because the unsuccessful memoization is replaced with actual computation, which in turn allows the succeeding layers to find more memoization opportunities. 

\begin{table}
\caption{Improvement with selective memoization.}
\label{selective_memoization}
\vspace{-10pt}
\scalebox{0.65}{
\begin{tabular}{cccc}
\hline
Model                   & Batch size & Inference time reduction & Memoization rate diff \\ \hline
\multirow{3}{*}{BERT}    & 1          & 5.0\%    & -6.8\%                \\
                         & 32         & 3.6\%    & -11.4\%               \\
                         & 64         & 8.7\%    & -9.5\%                \\ \hline
\multirow{3}{*}{RoBERTa} & 1          & 8.1\%    & -0.54\%               \\
                         & 32         & 12.3\%   & -0.78\%               \\
                         & 64         & 5.3\%    & -0.91\%               \\ \hline
\multirow{3}{*}{DeBERTa} & 1          & 3.0\%    & -1.1\%                \\
                         & 32         & 6.1\%    & -2.2\%                \\
                         & 64         & 4.2\%    & -2.2\%                \\ \hline
\multirow{3}{*}{GPT-2}   & 1          & 6.2\%    & 2.6\%                 \\
                         & 32         & 6.3\%    & 6.3\%                 \\
                         & 64         & 3.0\%    & 1.8\%                 \\ \hline
\end{tabular}
}
\vspace{-20pt}
\end{table}

\subsection{Scalability Analysis}\label{eval.Scalability}
\begin{figure}[t]
\centering
\subfloat[Sequence Length=16]{\includegraphics[width=0.23\textwidth]{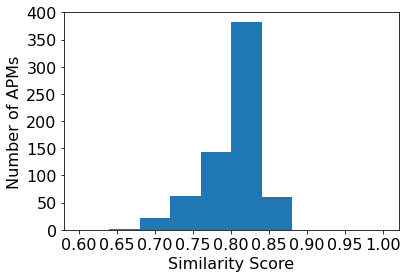}\label{fig:sub1}} 
\subfloat[Sequence Length=32]{\includegraphics[width=0.23\textwidth]{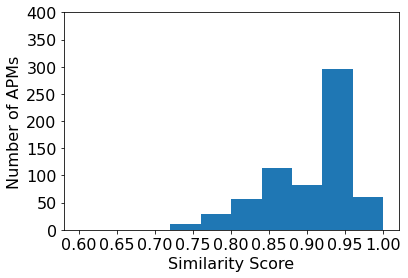}\label{fig:sub2}}\hskip1ex 
\subfloat[Sequence Length=64]{\includegraphics[width=0.23\textwidth]{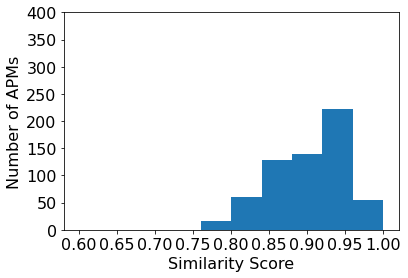}\label{fig:sub3}} 
\subfloat[Sequence Length=128]{\includegraphics[width=0.23\textwidth]{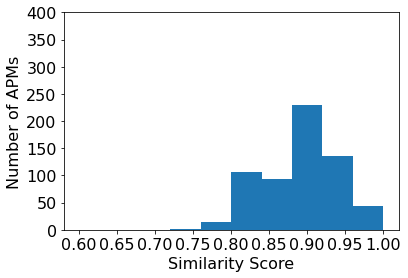}\label{fig:sub4}}
\vspace{-10pt}
\caption{Distribution of similarity scores in BERT with different input sequence lengths.} 
\vspace{-12pt}
\label{input_length}
\end{figure}

\begin{figure}[t]
    \centering
    \includegraphics[width=0.42\textwidth]{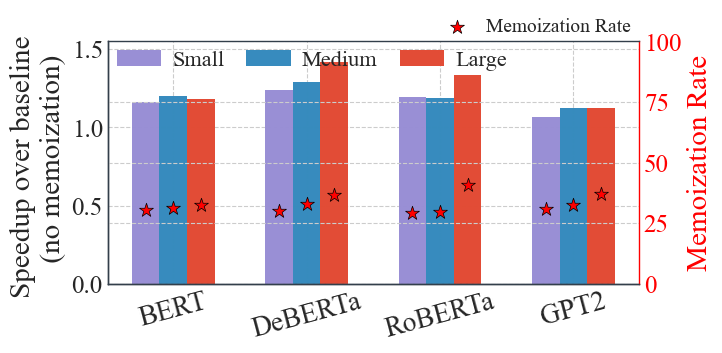}
    \vspace{-10pt}
    \caption{Memoization with different database sizes.}
    \label{database_size_sensitivity}
\end{figure}

\textbf{Increasing attention database size.} We increase the attention database size and study its impact on the memoization rate and inference time. In our evaluation, the memoization level is set as moderate and the batch size is set as 32. We use three different sizes of the attention database: for GPT-2, the attention database is built with the input sequence length of 1024 and the three database sizes are listed in Table~\ref{evaluation_detail}; for the other three models, the database is built with the input sequence length of 512 and the three database sizes are listed in Table~\ref{evaluation_detail} too. Figure~\ref{database_size_sensitivity} shows the results.  When we double the number of APMs in the database (almost doubling the database size from 630GB to 1.25TB for GPT-2 and from 575GB to 1.13TB for the other three models), the memoization rate increases by 2.3\%, 6.5\%, 11.5\%, and 5\% for BERT, DeBERTa, RoBERTa, and GPT respectively. As a result, we see 1.8\%, 18\%, 14\%, and 5.6\% reduction in inference time respectively. This result shows that increasing the attention database size provides more opportunities to successfully apply memoization, which in turn leads to better performance. In addition, doubling the database size causes less than 1\% variance in the search time because of efficient HNSW.

\textbf{Increasing input sequence length.} 
The input sequence length can impact the memoization performance because longer sequences might have more similarities across them. To evaluate the impact of the sequence length, we use 10,000 sequences from the SST2 dataset and develop an attention database while running BERT. All input sequences are bounded by a specific size, shown in Figure~\ref{input_length}. The figure shows the distribution of similarity scores, using the similar method as in Figure~\ref{fig:similarities}. Using a longer length (e.g., 128), the average similarity score is 0.87, while using a shorter length (e.g., 16), it is 0.79. Using a longer input sequence is helpful to introduce more similarities in self-attention.


\subsection{Complement to Sparsity-based Method}
The sparsity-based method applies parameter-pruning 
to 
transformer models to improve 
performance \cite{wang2021spatten, zafrir2021prune, cui2019fine, gordon2020compressing, xu2021rethinking, kim2022learned}. Compared with the memoization, 
the sparsity-based method pays extra efforts to change model topology and extensively prune-and-test the model 
to avoid accuracy loss. Nevertheless, \name can complement the sparsity-based method to reduce inference time.

\begin{figure}[!t]
    \centering
    \includegraphics[width=0.42\textwidth]{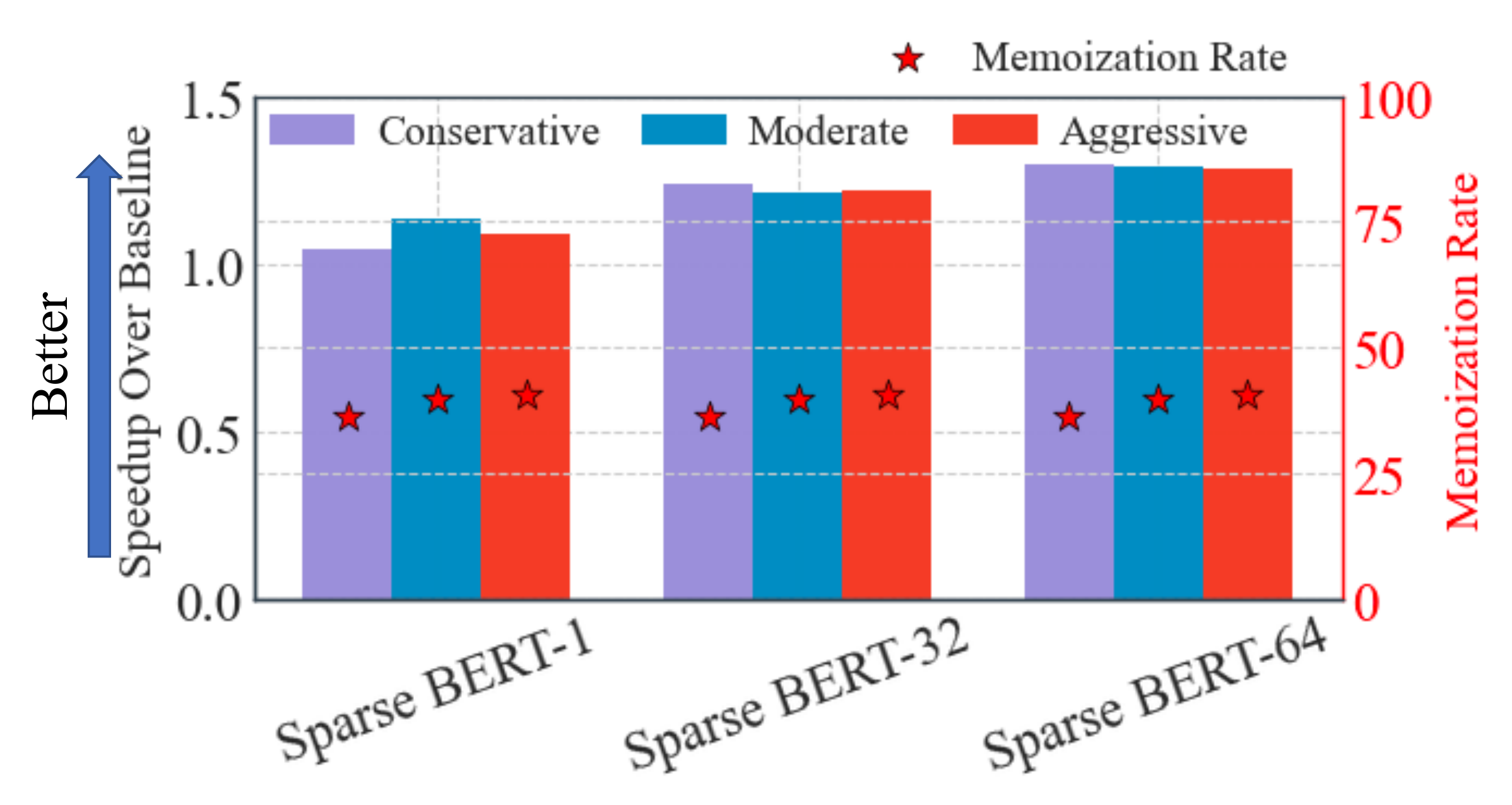}
    \vspace{-10pt}
    \caption{Applying \name to sparse models.}
    \label{comparsion_with_sparsity}
    \vspace{-10pt}
\end{figure}

To evaluate the effectiveness of \name with the sparsity-based method, we apply \name 
to three state-of-the-art pruned transformer models~\cite{zafrir2021prune} (with over 85\% parameters pruned). 
Figure~\ref{comparsion_with_sparsity} and Table~\ref{comparsion_acc_with_sparsity} show the results. 
When using \name with the conservative-level memoization, the three sparse models 
accelerate inference by 19\% on average with less than 1\% loss in accuracy.

\begin{table}[!t]
\caption{Accuracy after applying \name to sparse models}
\vspace{-10pt}
\scalebox{0.90}{
\begin{tabular}{ccccc}
\hline
Batch Size               & baseline         & Conservative & Moderate & Aggressive \\ \hline
1             & 90.31\%            & 89.69\%        & 88.12\%    & 82.97\%      \\
32            & 90.31\%            & 89.69\%        & 88.12\%    & 82.97\%      \\
64            & 90.31\%            & 89.69\%        & 88.12\%    & 82.97\%      \\
Average Diff. & N/A & -0.62\%        & -2.18\%    & -7.34\%      \\ \hline
\end{tabular}
}
\vspace{-10pt}
\label{comparsion_acc_with_sparsity}
\end{table}

\subsection{Potential with Large Language Model}
\label{sec:llm_potential}
The large models (such as LLaMA and BLOOM) recently drew many attentions because of their capability to perform new tasks based on a few demonstrations or natural language instructions~\cite{Bloom+:GCS62}. Those models are often based on the transformers using the self-attention mechanism. We study the potential of using memoization with a large language model, LLaMA from Meta, as an example.

\textbf{Benefits of using CPU over GPU for model  inference.} LLaMA with 65-billion parameters consumes 147GB memory for the model inference. We use the Oracle cloud for evaluation. In particular, we use four GPU instances, and each has two NVIDIA A10 GPUs (each GPU has 24GB memory, and there are eight GPUs in total). Within an instance, GPUs are connected with NVLink with 600 GB/s bandwidth; across instances, the instances are connected with Ethernet over Infiniband (EoIB) with 9.79 GB/s bandwidth. For the LLaMA inference, we also use the CPU instance equipped with Intel Xeon(R) Platinum 8358 CPU (2.60GHz) with 64 cores and 1024 GB memory. A single CPU instance has large enough memory for the LLaMA inference. We use the example prompt in the LLaMA repo~\cite{lambdalabsml} to generate 2,000 output tokens.

Table~\ref{tab:llm_potential} summarizes the results. In the table, the hardware acquisition cost for a GPU instance includes the cost for the GPU, CPU (an Intel Platinum 8358 CPU) and 128GB memory; The hardware acquisition cost for a CPU instance includes the cost for the CPU (an Intel Platinum 8358 CPU) and one TB memory. The table shows that using multiple CPU instances can perform better than using multiple GPU instances by 9\% for the LLaMA inference, and the acquisition cost and Oracle cloud cost~\cite{oracle} of using CPUs are 1.29$\times$ and 1.8$\times$ cheaper respectively than using GPUs.

\begin{figure}[!t]
\centering
\subfloat[Similarity in Layer 0]{\includegraphics[width=0.23\textwidth]{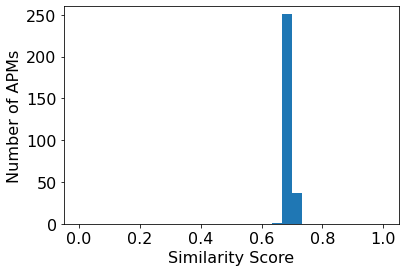}\label{fig:llama_0}} 
\subfloat[Similarity in Layer 15]{\includegraphics[width=0.23\textwidth]{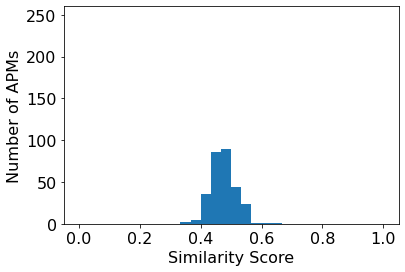}\label{fig:llama_15}}
\vspace{-12pt}
\caption{Distribution of similarity scores in LLaMA.}
\vspace{-12pt}
\label{fig:llama_similarity}
\end{figure}

\textbf{Memoization potential.}  Figure~\ref{fig:llama_similarity} shows the distribution of similarity scores in two representative layers (0th and 15th) in a 32-layer 7-billion-parameter LLaMA. We collect the similarity scores using a similar method as in Section~\ref{sec:motivation}. In particular, we use 2,048 as the input sequence length and collect 2,778 APMs from the training set of SST2 as the attention database. Then we find the most similar APMs in the database for 288 input sequences from the testing set of SST2. Figure~\ref{fig:llama_similarity} shows that in Layer 0, all APMs have high similarity scores (larger than 0.6), showing great potential to apply memoization; Layer 15 has less potential than Layer 0, but there are still 24\% APMs with similarity scores larger than 0.5.

\begin{table}[!t]
\footnotesize
\caption{The benefits of using CPU for large model inference. The numbers in parenthesis are the cost saving, compared with using 4 GPU instances.}
\vspace{-10pt}
\begin{tabular}{|c|c|c|c|}
\hline
\textbf{}                                                                   & \textbf{\begin{tabular}[c]{@{}c@{}}4 GPU instances \\ (8 GPUs in total)\end{tabular}} & \textbf{1 CPU instance} & \textbf{6 CPU instances} \\ \hline
\begin{tabular}[c]{@{}c@{}}Performance \\ (tokens/s)\end{tabular} & 5.54                                                                                    & 1.01                      &   6.06    \\ \hline
HW Acq. cost (\$)                                               & 61,200                                                                                      & 7,900 ($\downarrow$ 7.7x)                       & 47,400 ($\downarrow$ 1.29x 
)                        \\ \hline
Cloud cost (\$)                                                             & 1.6                                                                                     & 0.88 ($\downarrow$ 1.8x)                       & 0.88 ($\downarrow$ 1.8x)                        \\ \hline
\end{tabular}
\label{tab:llm_potential}
\vspace{-18pt}
\end{table}


\section{Conclusions}
The emerging big memory system brings new performance optimization opportunities. In this paper, we accelerate self-attention (a common model component in transformers) on big memory systems based on memoization. Our work is based on a unique observation that there is similarity in self-attention computation across transformer inferences. We introduce a framework \name using embedding, memory mapping, and performance modeling to make memoization feasible to accelerate transformer inferences. \name brings large performance improvement by 22\% on average (up to 68\%), compared with no memoization.


\bibliographystyle{plain}
\bibliography{references,li}

\clearpage

\end{document}